%% file: main.tex
\RequirePackage{fix-cm}
\documentclass[natbib,smallextended]{svjour3} 

\smartqed  

\usepackage{graphicx}%
\usepackage{aps-bibstyle}  
\usepackage{pdflscape}

\newcommand{\Sersic}{S\'{e}rsic }
 \journalname{ISSI Book on TDEs}

\usepackage{ref_macros} 
\usepackage{graphicx}
\usepackage{amssymb}
\usepackage{xspace}
\usepackage[usenames,dvipsnames,svgnames,x11names,table]{xcolor}

\definecolor{Purple}{rgb}{0.5,0,0.5}
\definecolor{Orange}{rgb}{1,0.5,0}

\graphicspath{{./}{figures/}}

\begin{document}

\title{The Host Galaxies of Tidal Disruption Events}



\author{K. Decker French \and Thomas Wevers \and Jamie Law-Smith \and Or Graur \and Ann I. Zabludoff
}


\institute{K. Decker French \at Observatories of the Carnegie Institute for Science, 813 Santa Barbara St., Pasadena CA 91101, USA \\
             \email{kfrench@carnegiescience.edu}\\
             ORCID: 0000-0002-4235-7337 \and
            Thomas Wevers \at Institute of Astronomy, University of Cambridge, Madingley Road, Cambridge CB3 0HA, United Kingdom \\
             \email{tw@ast.cam.ac.uk}\\
             ORCID: 0000-0002-4043-9400
\and
Jamie Law-Smith \at Department of Astronomy and Astrophysics, University of California, Santa Cruz, CA 95064, USA \\
\email{lawsmith@ucsc.edu}\\
ORCID: 0000-0001-8825-4790
             \and
            Or Graur \at Harvard-Smithsonian Center for Astrophysics, 60 Garden Street, MA 02138, USA \\
            \email{or.graur@cfa.harvard.edu}  \\
            ORCID: 0000-0002-4391-6137
            \and
            Ann Zabludoff \at Steward Observatory, University of Arizona, 933 N Cherry Ave, Tucson, AZ 85721, USA \\
            \email{aiz@email.arizona.edu} \\
            ORCID: 0000-0001-6047-8469
}

\date{Received: date / Accepted: date}

\maketitle

\begin{abstract}
Recent studies of Tidal Disruption Events (TDEs) have revealed unexpected correlations between the TDE rate and the large-scale properties of the host galaxies. In this review, we present the host galaxy properties of all TDE candidates known to date and quantify their distributions. We consider throughout the differences between observationally-identified types of TDEs and differences from spectroscopic control samples of galaxies. We focus here on the black hole and stellar masses of TDE host galaxies, their star formation histories and stellar populations, the concentration and morphology of the optical light, the presence of AGN activity, and the extra-galactic environment of the TDE hosts. We summarize the state of several possible explanations for the links between the TDE rate and host galaxy type. We present estimates of the TDE rate for different host galaxy types and quantify the degree to which rate enhancement in some types results in rate suppression in others. We discuss the possibilities for using TDE host galaxies to assist in identifying TDEs in upcoming large transient surveys and possibilities for TDE observations to be used to study their host galaxies.

\end{abstract}

\section{Introduction}
\label{sec:intro}
Tidal Disruption Events (TDEs) are observed when a star passes close enough to a supermassive black hole (SMBH) to be disrupted and torn apart by tidal forces. The rate of TDEs and the properties of the stars and SMBHs involved depend on the nuclear conditions of the host galaxies. 

The mass of the SMBH will affect whether TDEs are observed and which stars can be tidally disrupted outside the event horizon \citep[e.g.,][]{Hills1975,Rees1988,MacLeod2012,Law-Smith2017a}. Because the tidal radius scales as M$_{\rm BH}^{1/3}$, while the gravitational radius scales linearly with M$_{\rm BH}$, stars will be swallowed whole if the SMBH is larger than the so-called Hills mass \citep{Hills1975}, which is $\approx$10$^{8}$ M$_{\odot}$ for a non spinning black hole and a Solar type star. The Hills mass also depends on the SMBH spin, such that a faster spinning SMBH can disrupt less massive stars at a given M$_{\rm BH}$ \citep{Kesden2012}. The mass of the black hole is therefore an important parameter of a TDE. It can be estimated from observations because it is closely correlated with the mass and velocity dispersion of the galaxy's stellar bulge \citep{Magorrian1998, Ferrarese2000, Gultekin2009, Kormendy2013, McConnell2013}. The mass of the stars available to be disrupted depends on the recent star formation history of the galaxy and the initial mass function. The mass of the disrupted star is much harder to infer from observations, although it has been argued that it leaves imprints on the UV/optical lightcurve \citep{Lodato2009, Guillochon2015b, Mockler2018}.

Stars will be perturbed in their orbits to pass within the tidal radius depending on the distribution function of the stars in that galaxy \citep{Magorrian1999}. The parameter space of stars that can be tidally disrupted, called the loss cone, is thought to be re-filled mainly through two-body interactions, although other mechanisms may also play a non negligible role (see Stone et al. 2020, ISSI review). The stellar density profile in the vicinity of the SMBH and any deviations from an isotropic velocity / velocity dispersion field will hence affect the TDE rate \citep{Magorrian1999, Merritt2004, Stone2018}. 
As most TDEs are thought to be sourced from within the gravitational radius of influence of the black hole (\citealt{Stone2016b}; this is typically 0.1-10 pc for galaxies with M$_{\rm BH} \sim 10^{6-8} M_\odot$), the galaxy properties at these scales are likely to be most important in setting the TDE rate. However, the conditions in this region will be affected by the evolution and merger history of the galaxy as a whole, and may therefore be correlated with larger scale galaxy properties.

Large transient surveys such as the Palomar Transient Factory (PTF; \citealt{Law2009,Rau2009}), Pan-STARRS \citep{Chambers2016}, the All Sky Automated Survey for SuperNovae (ASASSN; \citealt{Shappee2014}), and the Zwicky Transient Facility (ZTF; \citealt{Bellm2019}) in the optical, as well as the Roentgen Satellite (ROSAT) and the X-ray
Multi-Mirror telescope (XMM; \citealt{Jansen2001}) in X-rays, and \textit{Swift} in gamma rays have enabled the detections of tens of TDEs, providing a sample large enough to study population properties. In addition to the TDE properties themselves, these new samples of TDEs also allow us to study trends in their host galaxy properties. \citet{Arcavi2014} studied the host galaxies of seven UV/optical bright TDEs with broad H/He emission lines, and found many of the hosts showed E+A, or post-starburst, spectra. Such post-starburst spectra are characterized by a lack of strong emission lines, indicating low current star formation rates, but with strong Balmer absorption, indicating a recent burst of star formation (within the last $\sim$Gyr) that has now ended. Quiescent Balmer-strong galaxies, and the subset of post-starburst/E+A galaxies with less ambiguous star formation histories, are rare in the local universe, and yet are over-represented among TDE host galaxies \citep{French2016, French2017, Law-Smith2017, Graur2018}. 

The observed correlations between the pc-scale regions of stars which can be tidally disrupted and the kpc-scale star-formation histories and stellar concentrations are a puzzle, for which many possible solutions have been proposed. Here, we review the known host galaxy properties of TDEs observed to date and with published or archival host galaxy spectra in \S\ref{sec:knownproperties}. We discuss possible drivers for the host galaxy preference in \S\ref{sec:discussion} and implications for the TDE rates in \S\ref{sec:rates}. We discuss possibilities for using the host galaxy information in future surveys to find more TDEs in \S\ref{sec:surveys} and study galaxy properties in \S\ref{sec:galaxystudies} and conclude in \S\ref{sec:summary}.

\subsection{TDEs Included in This Review}
In this review, we have selected a list of TDEs to discuss from the sample compiled by \citet{Auchettl2017} of X-ray and optical/UV - bright TDEs, as this sample has been used for recent host galaxy studies \citep{Law-Smith2017, Wevers2017, Graur2018}. Given the focus of this chapter on the host galaxy properties, we only include TDEs for which a spectrum of the host galaxy has been published or is available from archival surveys. We have added three more recent TDEs for which host galaxy spectroscopy is available from before the TDE from SDSS and BOSS: AT2018dyk, AT2018bsi, and ASASSN18zj (aka AT2018hyz), as well as a new TDE with published host galaxy information \citep[PS18kh,][]{Holoien2018b}.

We note the important caveat that the classification of transient events as TDEs is complicated by the heterogeneous datasets obtained for each event. For the purposes of this review we aim to balance including a large enough sample to reflect the range of published literature in this field with giving preference to the most well-justified claims of observed TDEs. We thus preserve the classifications of \citet{Auchettl2017} for the X-ray and optical detected TDEs that rank the likelihood an event is a TDE based on the completeness of the data, and divide the data into a number of subsets. It is important to note that the host galaxy statistics may change depending on which subset of TDE candidates are used. We comment on the differences one obtains depending on the sample used throughout, though for some subclasses we are limited by small number statistics.

We list in Table \ref{tab:tde_info} the TDEs considered in this review. 

We divide the TDEs into two classes---X-ray bright and optical/UV bright---with several sub-categories. The X-ray bright TDEs are subdivided further into X-ray TDEs  \citep{Holoien2015, Levan2011, Saxton2017, Holoien2016}, likely X-ray TDEs \citep{Saxton2012, Esquej2007, Cenko2012, Maksym2010, Lin2015, Lin2017} , and possible X-ray TDEs \citep{Komossa1999b, Gezari2008, Grupe1999, Ho1995, Greiner2000, Maksym2014} as done by \citet{Auchettl2017}. TDEs with no known or observed X-ray emission are classed as optical/UV TDEs \citep{Brown2017, Blanchard2017, Tadhunter2017, Gezari2009, Chornock2014, Komossa2009, Wang2012, VanVelzen2011, Yang2013, Holoien2015, Arcavi2014, Blagorodnova2019, 2018ATel11953....1A, 2018ATel12035....1G, 2018ATel12198....1D}. TDEs requiring re-classification based on new X-ray data are re-classified (ASASSN-15oi, PS18kh; K. Auchettl, private communication). 

TDEs that exhibited coronal lines \citep{Komossa2009,Wang2012} or broad H/He lines \citep[e.g.,][]{Arcavi2014} are also indicated. Three X-ray bright TDEs (ASASSN-14li, ASASSN-15oi, and PS18kh) additionally had significant optical observations, including broad H/He lines in their spectra, and are categorized as noted in the text. D3-13 is classed as a possible X-ray TDE, and also had significant optical/UV flux, but did not show broad H/He lines. We note that these classes are based on observational distinctions, which may or may not reflect physically different phenomena. Some optical/UV TDEs may have produced significant X-ray flux which was missed because of a lack of simultaneous X-ray observations. Indeed, the optical to X-ray luminosity ratios show significant variation in the events so far detected in both X-ray and optical light, and the observations of ASASSN-15oi by \citet{Gezari2017} demonstrate that X-ray emission can even be delayed well past the peak of the optical light curve, and would have been likely missed for many optical TDEs\footnote{See e.g. \citet{Jonker2019} for more late-time X-ray detections of UV/optical TDEs; this article was posted to the arXiv during review of this article.}. Similarly, the coronal line detections may be a light echo from a previous TDE \citep{Komossa2008}, and the relation between these events and the others is still unclear. These classes represent those for which samples have been aggregated in the literature, and with adequate host galaxy localization and observations to study for the purposes of this chapter. We direct the reader to the other chapters in this review, especially those by Saxton et al., Arcavi et al., Zauderer et al., Alexander et al., and Zabludoff et al., (2020, ISSI review) 
for further discussion of TDE classification and the question of multiple TDE classes.

In particular, we note that the class of events including F01004 \citep{Tadhunter2017} may be a type of nuclear phenomenon other than a TDE. \citet{Trakhtenbrot2019} argue against the TDE interpretation of this event as it has significantly narrower He lines than other broad H/He line events and the presence of Bowen fluorescence lines. However, Bowen fluorescence lines have now been found in other TDEs with broader H/He lines \citep{Leloudas2019}, indicating the space for observed TDE features may be broader than expected. The optical features of TDEs are discussed further in Arcavi et al. (2020, ISSI review), and a comparison of observational properties of observed TDE candidates with the spectrum of possible ``imposters" is discussed further in Zabludoff et al. (2020, ISSI review).

We separate out 13 TDE host galaxies which are part of the SDSS main spectroscopic sample (indicated in Table \ref{tab:tde_info}) in some of the following analysis, as these host galaxies can be matched to the general galaxy population in a uniform way.

\begin{table*}
    \centering
    \begin{tabular}{l c c c c c c}
    \hline
    Name & R.A. & Dec & $z$ & Type & BL$^a$  & CL$^b$ \\
    \hline
    \input{tde_info.txt}
    \end{tabular}
    \caption{TDEs included in this review. $^a$ Broad H/He lines observed during TDE. $^b$ Coronal lines observed. $^c$ Host galaxies with SDSS spectroscopy. Right ascension and declination are for the host galaxies.}
    \label{tab:tde_info}
\end{table*}

\section{Known Host Galaxy Properties of all events}
\label{sec:knownproperties}

We consider here the host galaxy properties and trends of the TDE samples discussed above. We compare the stellar mass and black hole masses to expectations given the volume-corrected mass functions and expectations from an upper cutoff in the black hole mass from event horizon suppression (\S\ref{sec:stmass}). We also consider the stellar populations and inferred recent star formation histories of the TDE hosts, and discuss the observed enhancement in post-starburst and quiescent Balmer-strong galaxies (\S\ref{sec:sfh}). We discuss the morphologies and concentrations of the stellar light and observed trends toward higher central concentration on kpc scales in the TDE hosts (\S\ref{sec:conc}). The presence of on-going gas accretion and AGN activity in the TDE host galaxies, as well as possible biases against identifying TDEs in such host galaxies are also discussed (\S\ref{sec:agn}). We summarize the extragalactic environments of the TDE host galaxies, given the efforts to identify TDEs in galaxy clusters (\S\ref{sec:enviro}).

The redshift range of the host galaxies affects the extent to which they can be studied. Most of the TDEs discovered to date are at low redshift, such that many of the TDE host galaxies have data from the SDSS. 
The redshift of all of the TDEs considered in this review (see Table \ref{tab:tde_info}) ranges from 0.01 to 0.4. The median redshift is $z=0.08$, and the 50 percentile range is 0.05--0.15. This redshift range is necessarily biased by the surveys which have discovered TDEs so far. Future surveys, such as LSST, may find a larger sample of higher redshift TDEs, depending on how the intrinsic TDE rate changes with redshift. A study by \citet{Kochanek2016} predicts that the TDE rate will drop steeply with redshift between $z=0$ and $z=1$, based on the expected evolution of the host galaxy stellar populations, black hole masses, and merger rates. However, the rising fraction of post-starburst hosts with redshift \citep{Yan2009, Snyder2011, Wild2016} may act to counter this effect. The blue continuum of TDE emission may result in a negative $k$-correction \citep{Cenko2016}, which would result in a greater number of observed TDEs at higher redshift.

\subsection{Host Galaxy Stellar Mass and Black Hole Mass}
\label{sec:stmass}

\begin{table*}
\centering
\begin{tabular}{l c c c c }
\hline
Name & log(M$_\star$) &M$_g$ & $\sigma$ & log(M$_{\rm BH}$) \\
 & (M$_{\odot}$) & (mag) & (km s$^{-1}$) & (M$_{\odot}$) \\
\hline
\input{tde_table_21.txt}
\hline
\end{tabular}
\caption{Table of properties of known TDE host galaxies discussed in \S\ref{sec:stmass}.
M$_\star$ (stellar mass) and M$_g$ ($g$-band absolute magnitude) are calculated as in \citet{Wevers2019}. $\sigma$ is the measured velocity dispersion, which is used to derive the black hole mass using the M--$\sigma$ relation of \citet{Ferrarese2005}. Measurement uncertainties are given between parentheses. The uncertainties in M$_{\rm BH}$ are the linear addition of the measurement uncertainties and the scatter in the M--$\sigma$ relation. Because the scatter in the M$-\sigma$ relation is a systematic uncertainty and the uncertainty in the velocity dispersion measurements are statistical, we add them linearly.}.  $\dagger$ Broad line TDEs (see Table \ref{tab:tde_info}). $^*$ Coronal line TDEs (see Table \ref{tab:tde_info}).
\label{tab:tde_properties}
\end{table*}

The black hole mass is one of the fundamental parameters for TDE studies, as it sets both the energetics (e.g. peak luminosity, accretion efficiency) and the dynamics (e.g. orbital timescales, relativistic effects) of the disruption. While theoretical predictions \citep{Wang2004} suggest that TDEs should preferentially occur in the lowest mass galaxies still hosting SMBHs (10$^4$--10$^6$ M$_{\odot}$), the observed distribution (using a heterogeneous set of measurements) was observed to peak around 10$^{7}$ M$_{\odot}$ (e.g. \citealt{Stone2016b, Kochanek2016}).  More recently, \citet{Wevers2017} presented systematic measurements of black hole masses using the M-$\sigma$ relation for a sample of 12 optical TDEs, and found the peak in the TDE black hole mass distribution to be significantly lower, near 10$^6$ M$_{\odot}$, consistent with theoretical predictions (taking into account the uncertainty in the calibration of the M-$\sigma$ relation at the low mass end). In contrast to previous studies \citep[e.g.,][]{Stone2016b} which use scaling relations from photometric observations to infer black hole masses, \citet{Wevers2017} use spectroscopic observations of the bulge velocity dispersions. Black hole mass measurements from TDE light curves \citep{Mockler2018} are consistent with measurements from galactic properties given the uncertainties in each set of measurements, but the number of TDE light curves with well-measured rises and thus more accurate black hole mass measurements is still limited.

\citet{vanvelzen2018} uses the BH masses from \citet{Wevers2017} to infer the BH mass and luminosity functions of TDEs. Correcting for selection effects such as survey depth, cadence and area, they find that the TDE rate is constant with black hole (or galaxy stellar) mass over two orders of magnitude from $M_{\odot} = 10^{5.5} - 10^{7.5}$. Given the uncertainties, the observed black hole mass function of TDE hosts could be consistent with either the expected black hole mass function over this mass range, or with the slightly steeper trend expected given the scaling of the TDE rate with black hole mass. The dearth of BH masses $\geq$ 10$^8$ M$_{\odot}$ is consistent with the presence of BH event horizons, and the disappearance of the tidal radius for a main sequence 1$M_\odot$ star inside the event horizon. 

\begin{figure*}
\begin{center}
\includegraphics[width=0.6\textwidth]{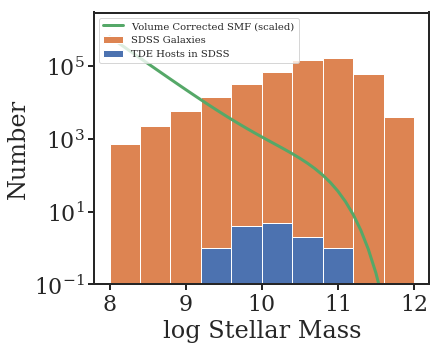}
\caption{Histogram of stellar masses for the SDSS main spectroscopic sample, with quality cuts as described in \citet{Law-Smith2017} (orange), TDE host galaxies from the SDSS main spectroscopic sample (blue), and the volume corrected stellar mass function from \citet{Baldry2012}. The TDE host galaxies are less massive than the typical SDSS galaxies, but with a typical stellar mass near M* (the turnover in the stellar mass function, measured by \citet{Baldry2012} to be $10^{10.66}$ M$_{\odot}$). This distribution is consistent with the TDE host galaxies being drawn from a volume limited sample of galaxies with stellar mass greater than $10^9$ M$_\odot$. }

\label{fig:stmass}
\end{center}
\end{figure*}

While black hole masses are difficult and time-consuming to measure, stellar masses can be more easily measured using galaxy luminosities and stellar population estimates. The stellar masses of the host galaxies are roughly correlated with the black hole masses, via the black hole mass -- bulge mass relation \citep[e.g.,][]{McConnell2013} and the correlation between galaxy stellar mass and bulge stellar mass \citep[e.g.,][]{Mendel2014}. For the host galaxies in the SDSS main spectroscopic sample, we plot a histogram of their stellar masses compared to the rest of the SDSS galaxies and the volume-corrected stellar mass function (SMF) in Figure~\ref{fig:stmass}. The TDE host galaxies are less massive than the typical SDSS galaxies, but with a typical stellar mass near M*. This distribution is consistent with the TDE host galaxies being drawn from a volume limited sample of galaxies with stellar mass greater than $10^9$ M$_\odot$.

Are the distributions of host galaxy stellar mass or black hole mass different for different classes of TDEs? There are several predictions in the literature, and this question depends on the details of how stars are disrupted and accreted, and the origin of the observed emission. The inverse dependence of the accretion disk temperature on black hole mass suggests X-ray TDEs should have lower black hole masses than optical TDEs \citep{Dai2015}, but if rapid circularization is required to produce X-rays, higher mass black holes may be expected to produce more X-ray emission \citep{Guillochon2015b}. Alternatively, if the difference between the classes is related to a viewing angle effect \citep{Dai2018}, no difference in the host galaxy properties would be expected.

\citet{Wevers2019} have measured the host galaxy absolute magnitudes of a large sample of optical and X-ray TDEs using SDSS and PS1 photometry. They used the kcorrect software \citep{Blanton2017} and the Petrosian or Kron magnitudes for SDSS and PS1 to estimate the host absolute magnitude as well as the galaxy stellar mass for a sample of 35 TDEs and TDE candidates. These values are presented in Table \ref{tab:tde_properties} for all sources in the current sample. Using different subdivisions (and a smaller sample) of host galaxies than the ones used here, \citet{Wevers2019} found that the host galaxy absolute magnitudes, stellar masses, and black hole masses for different TDE classes are consistent with being drawn from the same parent population. 

\begin{table}
    \centering
    \begin{tabular}{cccc}
\hline
Sample size & M$_{\star}$ & M$_{\rm g}$ & M$_{\rm BH}$ \\\hline
Optical &20 & 20 & 16\\
X-ray + likely X-ray & 10 & 10 & 7\\
Possible X-ray & 6 & 6 & 6 \\\hline
p-values & & & \\\hline
Optical - X-ray & 0.03 (0.09) & 0.02 (0.02)& 0.38 (0.42) \\
X-ray - pos. X-ray & 0.15 (0.05) & 0.03 (0.02) & 0.06 (0.10)\\\hline
\end{tabular}
    \caption{Summary of statistical comparison between samples for different host properties, including host galaxy stellar mass (M$_{\star}$, absolute $g$-band magnitude (M$_{\rm g}$) and the black hole mass (M$_{\rm BH}$). We give the relevant sample sizes for each parameter. We test the hypothesis that the respective samples are drawn from the same parent distribution. The p-values of an Anderson-Darling test are given, as well as the p-values for a Kolmogorov-Smirnov test (in parentheses); values below 0.05 suggest that we can reject the hypothesis at $>$95 $\%$ significance. We note that these conclusions differ from those by \citet{Wevers2019} due to the larger sample of TDEs considered here, as well as a different class division.}
    \label{tab:pvalues}
\end{table}

\begin{figure*}
\begin{center}
\includegraphics[width=0.8\textwidth]{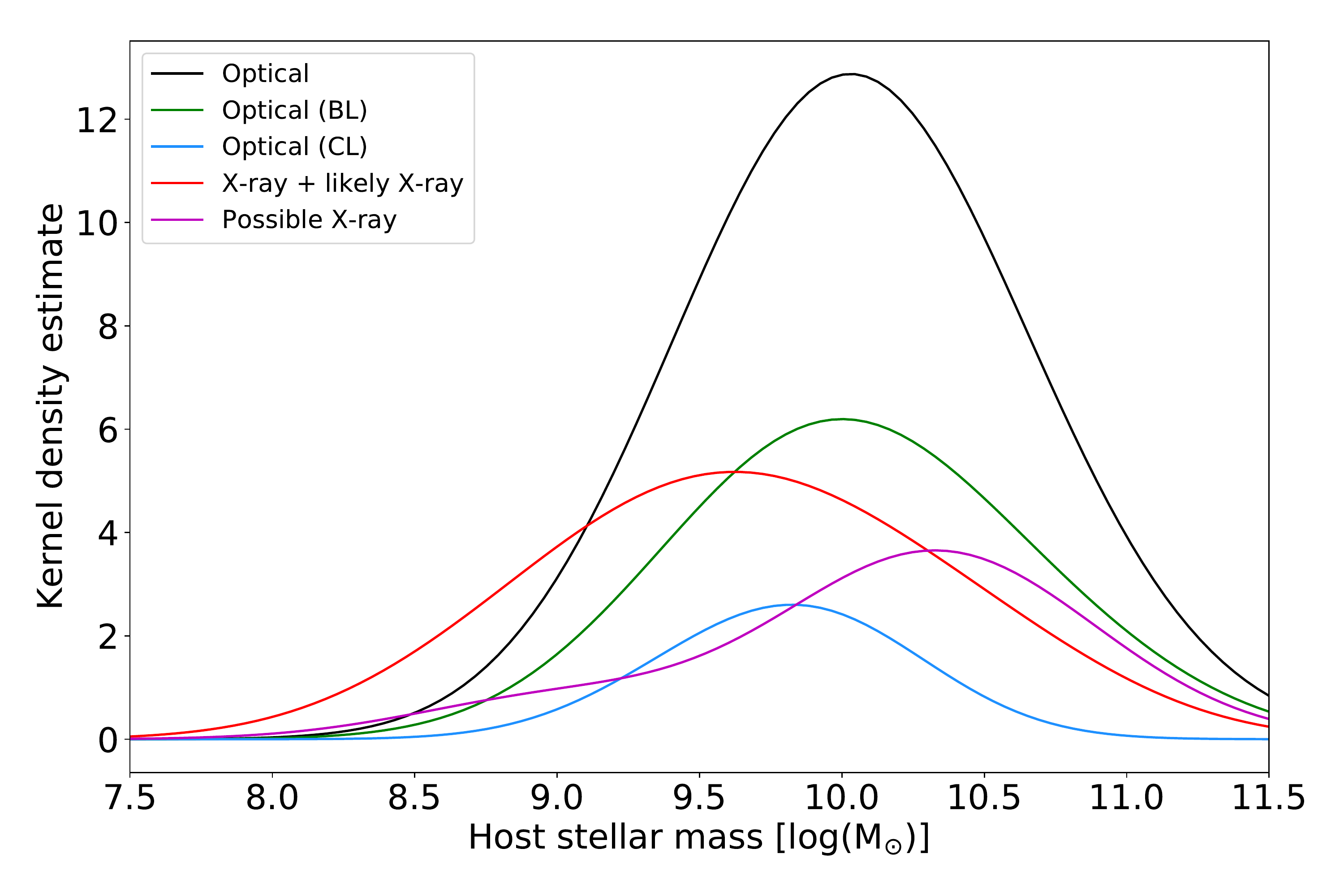}
\includegraphics[width=0.8\textwidth]{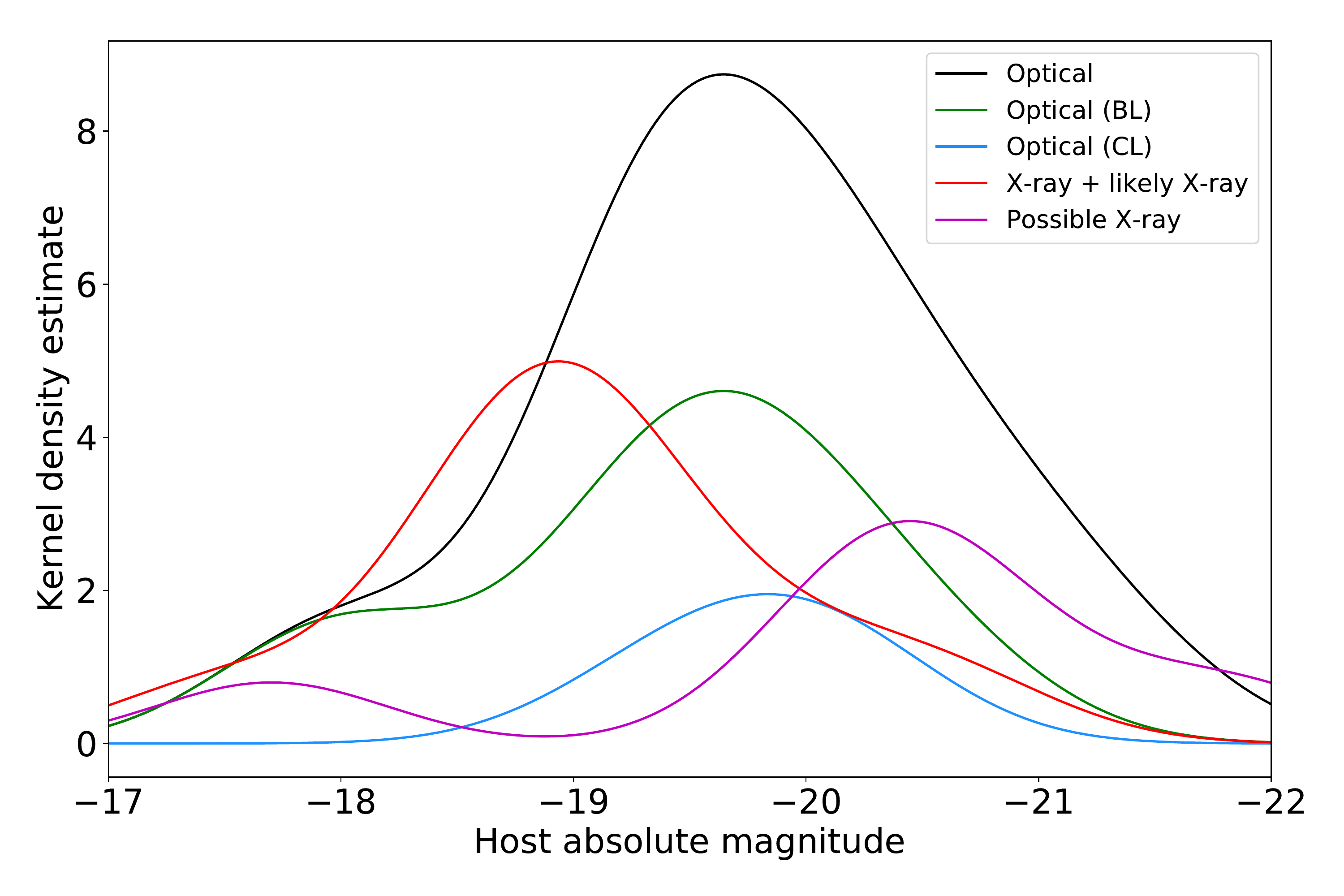}
\includegraphics[width=0.9\textwidth]{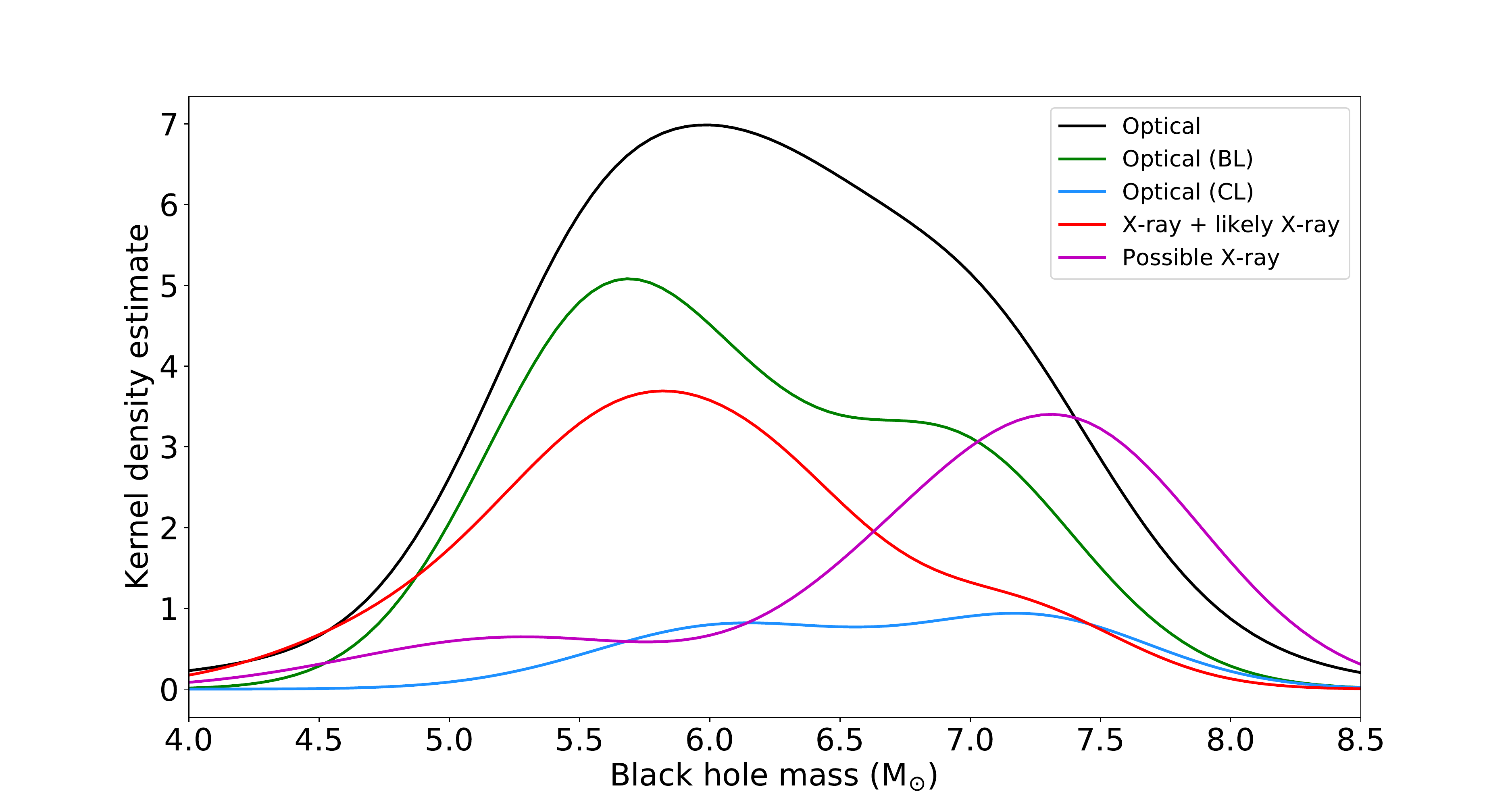}
\caption{\textbf{Top:} Kernel density estimate of TDE host galaxy stellar masses, sub-divided by TDE class. KDEs are calculated as in \citet{Wevers2019}, using Gaussians with width = 0.5 dex (the typical uncertainty on the host galaxy stellar mass measurements). Bulge to total light (B/T) corrections are applied (not individual measurements but a general prescription). The X-ray and optical/UV TDEs are consistent with being drawn from different parent samples, according to their Anderson-Darling statistic.  
\textbf{Middle:} Kernel density estimate of TDE host galaxy absolute magnitude, sub-divided by TDE class. KDEs are calculated as in \citet{Wevers2019}, using Gaussians with width = 0.5 dex (the typical uncertainty). The X-ray TDEs and optical TDEs are again consistent with being drawn from different parent samples. The possible X-ray TDE candidates are also distinct from the X-ray + likely X-ray samples as inferred from their Kolmogorov-Smirnov and Anderson-Darling tests.
\textbf{Bottom:} Kernel density estimate of the black hole masses of X-ray and UV/optical selected TDEs, sub-divided by TDE class. The kernel size includes the velocity dispersion measurement as well as the intrinsic M--$\sigma$ scatter. No statistically significant differences are observed between the TDE classes. Interestingly, while the X-ray TDEs prefer host galaxies with lower stellar mass and absolute magnitude than the optical/UV TDEs, this trend is not seen in black hole mass. More observations of black hole masses for various classes of TDEs, especially X-ray TDEs, are required to test whether this effect is physical.
}
\label{fig:kde}
\end{center}
\end{figure*}

The larger sample considered here allows us to repeat the analysis in \citet{Wevers2019} with more statistical power (Figure \ref{fig:kde}; Table \ref{tab:tde_properties}). We group the X-ray and likely X-ray hosts, the UV/optically discovered hosts and the possible X-ray hosts and perform pairwise Kolmogorov-Smirnov (KS) and Anderson-Darling (AD) tests for these 3 samples. For the X-ray and optical samples, we find that for both host galaxy stellar mass and absolute magnitude the hypothesis that they are drawn from the same parent distribution can be rejected. The KS and AD significance values are summarized in Table \ref{tab:pvalues}. The p-values for the X-ray and possible X-ray stellar mass comparison are higher, and we cannot reject the null hypothesis that they are drawn from the same parent distribution. For the latter, this could be due to the small size of the sample.
The properties of the possible X-ray sources suggest significant contamination by AGN, which favour higher mass (both stellar mass and M$_{\rm BH}$) and more luminous host galaxies. The difference with the results in \citet{Wevers2019} can be explained by i) the larger sample considered here and ii) the different sample subdivision. In particular, the soft X-ray sample in \citet{Wevers2019} consists of 6 likely and 6 possible X-ray TDE hosts, and the sources ASASSN--14li and ASASSN--15oi are considered as optical events.

\citet{Wevers2019} also presented velocity dispersion measurements of an additional 19 TDE candidates, yielding a sample of 29 homogeneously measured black hole masses\footnote{We note that the sample presented by \citet{Wevers2019} contains a smaller but not completely overlapping sample to that presented in this review, due to differences in the TDE selection and differences in available data.}. Figure \ref{fig:kde} shows a kernel density estimate (KDE) of the black hole mass distribution, divided by type. The KDE was calculated by representing each black hole mass estimate with a Gaussian function with a full width at half maximum (FWHM) equal to the measurement uncertainty (including both the velocity dispersion uncertainty and the scatter in the M--$\sigma$ relation), and then summing over the respective samples. 

Using statistical tests to compare the distributions between X-ray and optical samples, we find no significant differences between the black hole mass distributions. This supports the idea that the apparent dichotomy between optical and X-ray selected TDEs could be related to (for example) viewing angle or geometry \citep{Watarai2005,Coughlin2014,Dai2015,Metzger2016,Roth2016,Dai2018}, and that these events intrinsically belong to the same class. This is also supported by observations of UV/optical TDEs with deep X-ray upper limits: the detection of Bowen fluorescence lines in optical spectra implies that an ionizing (X-ray) radiation field exists, although no X-rays are actually observed \citep{Leloudas2019}. UV emission is not sufficient to excite the Bowen fluorescence lines for the one TDE (AT2018dyb) for which measurements are available. Emission line measurements presented by \citet{Leloudas2019} show that the Wien tail of the UV blackbody responsible for the UV/optical radiation is insufficient (by $\sim$6 orders of magnitude) to explain the observed line fluxes. This suggests that the X-ray source is completely obscured along the line of sight.

While the black hole mass distribution is very similar, the host galaxy stellar mass and absolute magnitude are significantly lower for the X-ray sample, as we can reject the null hypothesis of a common parent sample at high significance. However, we caution that this effect could be due to the lesser number of TDE host galaxies with black hole mass measurements (Table \ref{tab:pvalues}). A larger sample of robust X-ray TDEs is required to draw robust statistical conclusions, and test whether optical/UV TDEs might have smaller black hole masses for their stellar masses (or whether X-ray TDEs have larger black hole masses for the stellar masses).

If the observed trends in stellar mass and absolute magnitude are driven by differences in the black hole mass distribution between the X-ray and optical samples, this suggests that smaller black holes have higher temperature accretion disks with higher X-ray luminosities, or a combination of this effect and a viewing angle effect are acting. Another possibility is that selection biases from the very different identification methods of TDEs in the optical vs. X-ray could lead to this effect. The light curve duration for TDEs may vary with black hole mass in different ways for X-ray compared to optical emission. Wen et al. (in prep) find lower mass black holes to have longer duration super-eddington plateaus in their predicted X-ray light curves. \citet{Lin2018} have found one such example of a long duration X-ray light curve from an event around a small SMBH. If it were the case that smaller black holes have longer-duration light curves in the X-ray compared to the optical, coarser cadence surveys in the X-ray would be biased against detecting TDEs in more massive black holes. Differences between TDEs found in optical vs. X-ray surveys will need to be studied further in the era of eROSITA and LSST. 

\citet{Wevers2019} also consider a class of hard X-ray selected TDE candidates (which are not included in the sample discussed here), finding these host galaxies to have significantly different black hole mass distributions, as well as absolute magnitude and stellar mass distributions. However, these conclusions are based on a sample of 5 hard X-ray TDE candidate host galaxies, and a larger sample is needed to confirm these findings and understand the cause of these potential differences.

\subsection{Star Formation Rates, Star Formation Histories, and Stellar Populations}
\label{sec:sfh}

\begin{table*}
\centering
\begin{tabular}{l c c c c c c c c c}
\hline
Name & H$\alpha$ & $\sigma$(H$\alpha$) & Lick H$\delta_A$ & $\sigma$(Lick H$\delta_A$) & SFR & Type \\
 & [\AA] & [\AA] &[\AA] &[\AA] & M$_\odot$ yr$^{-1}$ & \\
\hline
\input{tde_table_22.txt}
\end{tabular}
\caption{Table of properties of known TDE host galaxies discussed in \S\ref{sec:sfh}. Negative values indicate emission and positive values indicate absorption for the H$\alpha$ and H$\delta$ measurements. SFRs are only calculated when SDSS MPA-JHU catalogue data are available. To be classified as quiescent, quiescent Balmer-strong (QBS), or post-starburst (PSB), galaxies are required to have H$\alpha$ EW $>$ -3 \AA. Galaxies with stronger H$\alpha$ emission are labeled as star-forming (SF). Galaxies with strong Balmer absorption (Lick H$\delta_{\rm A} >$ 1.31\,\AA) are classified as QBS, and a subset of these are classified as PSB if they meet the threshold of H$\delta_{\rm A}$ $-$ $\sigma$(H$\delta_{\rm A}$) $>$ 4\,\AA. 5/41 (12\%) host galaxies are post-starburst galaxies and 13/41 (32\%) are either QBS or PSB. Of the 4 X-ray TDEs, 3 (75\%) are QBS and 1 (25\%) is PSB. Of the 15 broad H/He line TDEs, 9 (60\%) are QBS and 5 (33\%) are PSB. We consider the effects of these choices of dividing criteria in \S\ref{sec:sfh}.  $\dagger$ Broad line TDEs (see Table \ref{tab:tde_info}). $^*$ Coronal line TDEs (see Table \ref{tab:tde_info}).}
\label{tab:tde_sfh}
\end{table*}

Next, we consider the current star formation rates, the past star formation history, and the stellar populations of the TDE host galaxies. While the stellar populations in the nucleus may be far removed from those of the bulk of the host galaxy, the galaxy-wide stellar populations trace the formation and evolution of the host galaxy and are closely tied to the morphologies, kinematics, interstellar medium properties, and merger histories of the host galaxies.

Star formation rates (SFRs) for host galaxies are calculated using various tracers of short-lived massive O and B stars. While many SFR tracers from the UV to the radio work well for galaxies with significant star formation, these tracers can be heavily biased in galaxies with rapidly changing SFRs (as many of the TDE host galaxies are thought to have), in dusty galaxies, or in galaxies influenced by AGN activity. 

\begin{figure*}
\includegraphics[width=1.0\textwidth]{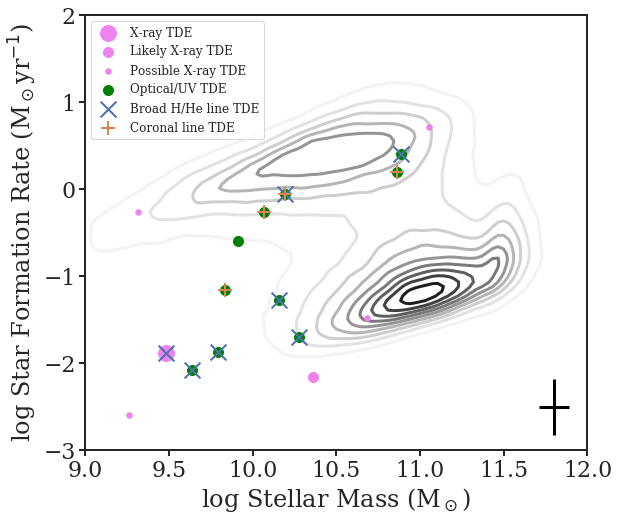}
\caption{Star formation rates vs. stellar masses for the SDSS main spectroscopic sample (grey contours) and the different classes of TDE hosts considered here. Stellar masses are from the MPA-JHU catalogue for all samples, as well as SFRs for the SDSS sample. We determine the TDE host SFRs based on H$\alpha$ fluxes as described in \S\ref{sec:sfh}. A characteristic error-bar is shown in the bottom right. While several of the TDE hosts are at the lower SFR edge of the ``main sequence" of star forming galaxies, most are quiescent, with low SFRs. We note that the SDSS sample has not been volume corrected; for a more detailed analysis of the stellar mass distributions of the TDE hosts compared to a volume-limited sample, see \S\ref{sec:stmass}. Thus, for the typical black hole masses of TDEs, the host galaxies lie in a relatively spare region on this diagram due to the magnitude-limited nature of the SDSS comparison sample used here.}
\label{fig:ms}
\end{figure*}

We consider here SFRs of TDE host galaxies derived using H$\alpha$ luminosities from the SDSS, and correct for extinction and aperture using the corrections from the MPA-JHU SDSS catalogues \citep{Brinchmann2004,Tremonti2004}. The SFRs from this catalogue use the D4000 break to estimate the SFRs for galaxies with non-star-forming emission line ratios, like those of many of the TDE hosts (see \S\ref{sec:agn}). While the D4000-sSFR correlation will lead to accurate SFRs on average for large samples, individual galaxies will have high uncertainties. Our use of H$\alpha$ here thus is more accurate for galaxies without strong AGN, but will be in general biased towards higher values for galaxies with additional H$\alpha$ emission from non-star-forming sources. 
We convert the H$\alpha$ luminosities to SFRs using $\eta = 5.4\times10^{-42}$ M$_\odot$ yr$^{-1}$/(ergs s$^{-1}$) \citep{Kennicutt2012}. 
We compare the TDE hosts to galaxies from the SDSS main spectroscopic survey in SFR--stellar mass space in Figure \ref{fig:ms}. While several of the TDE hosts are at the lower SFR edge of the ``main sequence" of star forming galaxies, most are quiescent, with low SFRs.

\begin{figure*}
\includegraphics[width=1.0\textwidth]{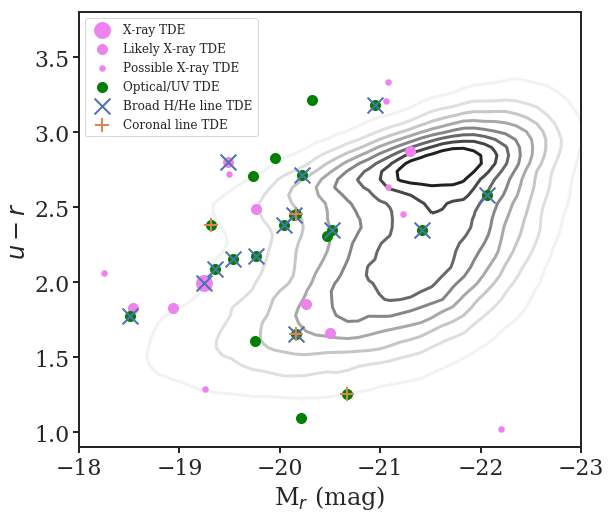}
\caption{Optical colour magnitude relation of TDE host galaxies and the SDSS main spectroscopic sample. All TDE hosts with SDSS photometry are plotted here. We note that the SDSS sample has not been volume corrected; for a more detailed analysis of the absolute magnitude distributions of the TDE hosts compared to a volume-limited sample, see \S\ref{sec:stmass}. The TDE hosts occupy a range of red, blue, and green-valley host galaxies. Because the colours are affected by both the current SFR and recent star formation history (SFH), we explore the physical interpretation of these quantities below.}
\label{fig:cmr}
\end{figure*}

We also consider the optical colours of the TDE hosts in the context of the colour magnitude relation. The $u-r$ colours and M$_r$ absolute magnitudes are plotted in Figure \ref{fig:cmr}. The TDE hosts occupy a range of red, blue, and green-valley host galaxies. Because the colours are affected by both the current SFR and recent star formation history (SFH), we explore the physical interpretation of these quantities below.

The quiescent SFRs paired with green colours of many of the TDE hosts is suggestive of an intermediate stellar population and a recent decline in the SFH of the host. Indeed, a large number of TDEs have been observed in E+A or post-starburst galaxies. The large number of post-starburst host galaxies was first observed by \citet{Arcavi2014} in a sample of UV/optical bright TDEs with broad H/He lines. The presence of even one post-starburst galaxy amongst the hosts would be unusual given the rarity of post-starburst galaxies. 

The over-representation of post-starburst galaxies in TDE hosts was then quantified by \citet{French2016}, who used tracers sensitive to recent star formation on different scales to assess the recent star formation history of the host galaxies.  One such comparison is to use the H$\alpha$ emission as a tracer for the current star formation on $\sim10$ Myr timescales and the Balmer absorption as a tracer for star formation on timescales of $\sim1$ Gyr. This method of using H$\alpha$ emission vs. Lick H$\delta_A$ has been used by many \citep{French2016, Law-Smith2017, Graur2018} to study the recent star formation histories of TDE host galaxies.

 There are many ways to quantify the current SFR in galaxies, as described above. Many methods that are sensitive to post-starburst galaxies use H$\alpha$ emission or O[II] emission if the red end of the rest-frame spectra are not available. Other methods allow for residual star formation using a PCA analysis \citep{Wild2010} or a BPT diagram analysis \citep{Alatalo2016}. \citet{French2016} require H$\alpha$ EW $<$ 3 \AA\ in emission in the rest frame to be considered quiescent. This corresponds to a specific SFR $\lesssim 1\times10^{-11}$ yr$^{-1}$, well below the main sequence of star-forming galaxies \citep[e.g.,][]{Elbaz2011}. The H$\alpha$ emission is also corrected for stellar Balmer absorption, which is significant for post-starburst quiescent Balmer-strong galaxies.

The moderate lifetime of A stars means that the presence of a large A star population is indicative of a burst of star formation within the last Gyr. A star spectra show strong Balmer absorption, which can be best traced using the H$\gamma$, H$\delta$, or H$\epsilon$ lines. \citet{French2016} use the Lick H$\delta_{\rm A}$ index and its uncertainty
$\sigma$(H$\delta_{\rm A})$, which is optimized for the stellar absorption from A stars \citep{Worthey1997}, has lower emission filling than H$\beta$, and smooth nearby continuum regions. The more bursty the SFH, i.e. the greater fraction of stellar mass produced over a shorter time, the higher H$\delta$ absorption will be. A stricter cut of H$\delta_{\rm A}$ $-$ $\sigma$(H$\delta_{\rm A}$) $>$ 4\,\AA\ will select galaxies with recent starbursts creating $>3$\% of their current stellar mass over 25--200 Myr (referred to as post-starburst galaxies throughout), and a weaker cut of H$\delta_{\rm A} >$ 1.31\,\AA\ (referred to as quiescent Balmer-strong galaxies throughout) will select galaxies with recent epochs of star formation which created $>0.1$\% of their current stellar mass over 25--1000 Myr \citep{French2017,French2018b}. 

\begin{figure*}
\includegraphics[width=0.5\textwidth]{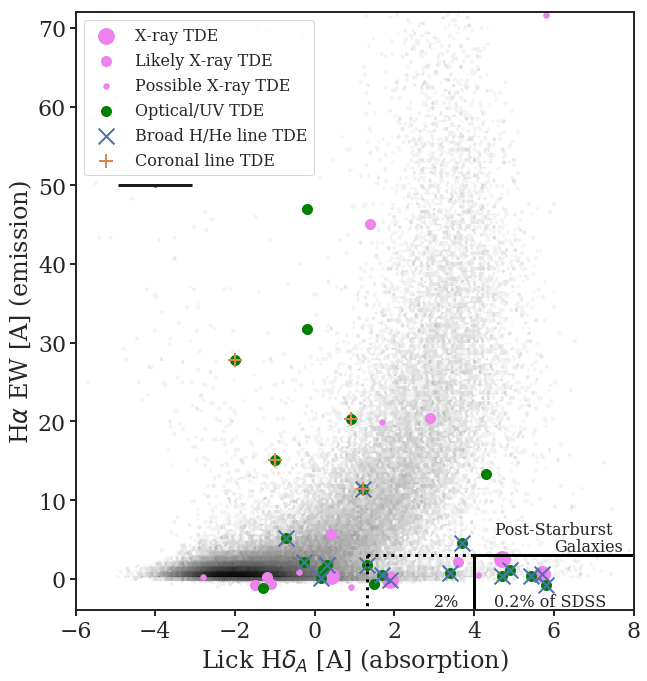}
\includegraphics[width=0.5\textwidth]{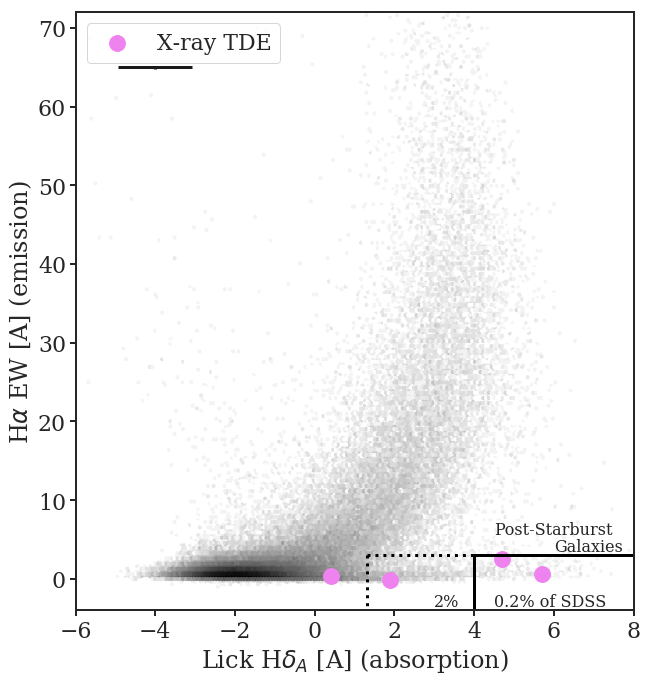}
\includegraphics[width=0.5\textwidth]{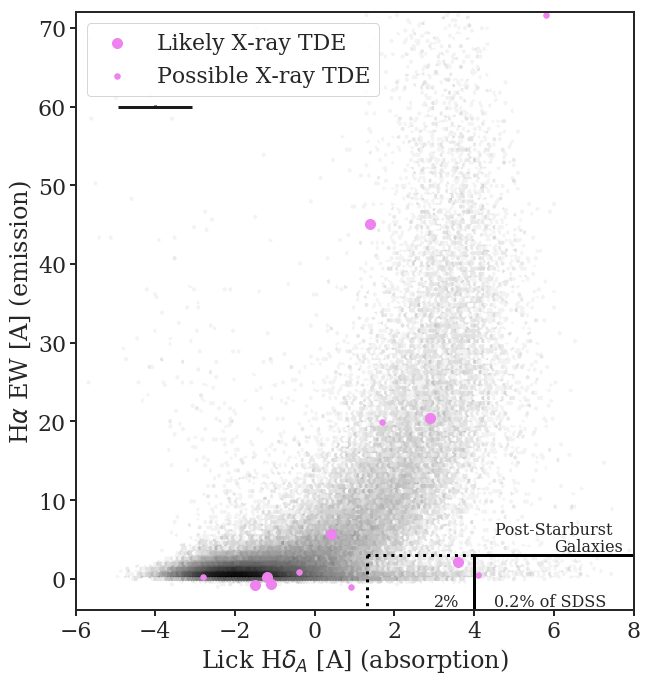}
\includegraphics[width=0.5\textwidth]{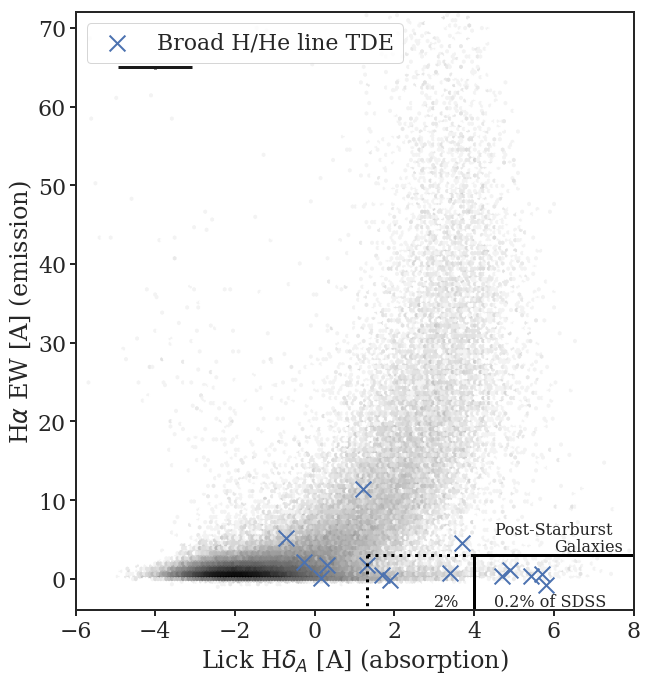}
\includegraphics[width=0.5\textwidth]{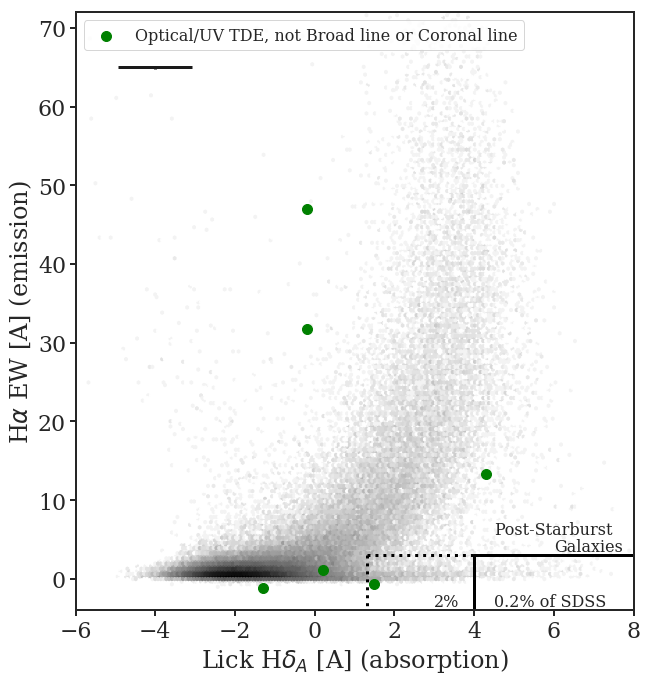}
\includegraphics[width=0.5\textwidth]{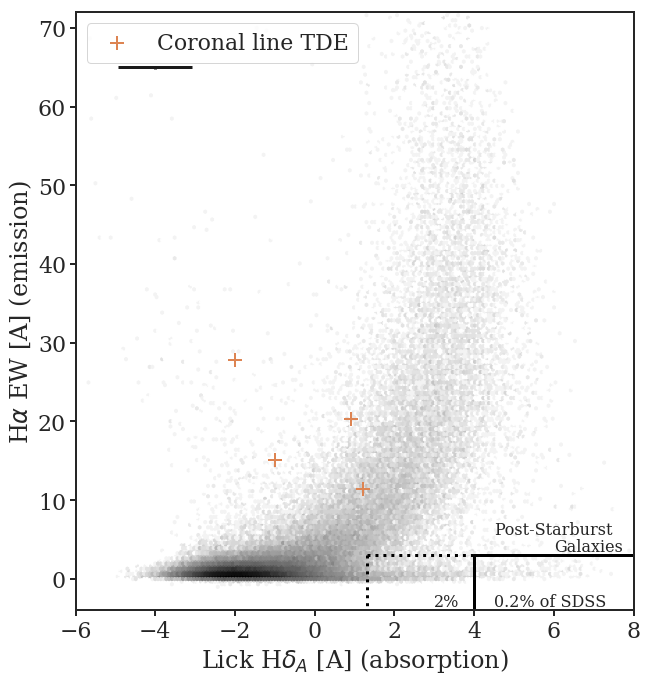}
\caption{Spectral indices tracing the recent star formation histories of TDE hosts and SDSS galaxies \citep[updated from][]{French2016, Graur2018}. We plot H$\alpha$ EW (sensitive to current star formation) vs. Lick H$\delta_{\rm A}$ absorption (sensitive to star formation over the past Gyr) for each galaxy. Galaxies with low current star formation, yet significant star formation over the past Gyr form the post-starburst/quiescent Balmer-strong ``spur" in the lower right. TDEs, especially the X-ray and broad H/He line classes, are over-represented among the post-starburst and quiescent Balmer-strong galaxies.}
\label{fig:hahd}
\end{figure*}

We plot these SFH tracers in Figure \ref{fig:hahd} for various subsamples of TDEs. The TDE host galaxies span a range of SFHs from star-forming galaxies, to quiescent galaxies which have been quiescent for at least the past Gyr, and galaxies which had significant star-formation within the last Gyr but are currently quiescent. This last category consists of ``post-starbust" or ``quiescent Balmer-strong" galaxies as defined above.

Several TDE host galaxies in the classes of the coronal line TDEs and optical/UV TDEs without coronal or broad lines have low H$\delta$ absorption for their H$\alpha$ emission compared to the rest of the galaxies in the SDSS spectroscopic sample. This may not be due to a physical difference between the host galaxies, and might instead be due to filling of the H$\delta$ line by residual TDE emission, as discussed by \citet{Graur2018}. Another possibility is contamination from the nearby Bowen fluorescence line NIII $\lambda$4100 \citep{Blagorodnova2018,Leloudas2019, Trakhtenbrot2019}. 

The over-representation of a galaxy type among the TDE host galaxies can be determined using its rate in the TDE host galaxies compared to its rate in a general galaxy sample. We describe here the analyses done by various groups, and summarize in Table \ref{tab:overenhancement}. \citet{French2016} find that 38\% of a sample of eight UV/optical H/He broad line TDE hosts meet a post-starburst selection criterion with a rate of only 0.2\% in the general galaxy population. Similarly, 75\% of the same TDE host galaxies meet a quiescent Balmer-strong selection criterion with a rate of 2.3\% in the general galaxy population. These rates imply overdensities of 33$^{+7}_{-11}\times$ in quiescent Balmer-strong galaxies and 190$^{+115}_{-100}\times$ in post-starburst galaxies. 

\citet{Graur2018} considered the over-enhancement rates for several additional categories of observed TDEs using a similar but slightly different parent galaxy sample and post-starburst/ quiescent Balmer-strong definitions. For an updated sample of UV/optical bright H/He broad line TDEs, the over-enhancement rates are 34$^{+22}_{-14}\times$ in quiescent Balmer-strong galaxies and 110$^{+80}_{-50}\times$ in post-starburst galaxies, consistent with the rate enhancements found by \citet{French2016}. 

The over-enhancement rates in post-starburst galaxies for the X-ray bright TDEs are weaker than for the UV/optical broad line TDEs, though this comparison is limited by small number statistics. For the set of X-ray TDEs, ``likely" X-ray TDEs, and ``possible" X-ray TDEs identified in \citet{Auchettl2017a}, \citet{Graur2018} find the over-enhancement rates to be 18$^{+13}_{-9}\times$ in quiescent Balmer-strong galaxies and 18$^{+22}_{-18}\times$ in post-starburst galaxies. These rates are higher once the ``possible" X-ray TDEs are excluded, many of which have ambiguous light curves and may be AGN flares. Considering only the X-ray and ``likely" X-ray TDEs, the over-enhancement rates are 23$^{+21}_{-13}\times$ in quiescent Balmer-strong galaxies and 29$^{+41}_{-29}\times$ in post-starburst galaxies. These rate enhancements for the post-starburst sample are driven by the one X-ray (including the ``likely" and ``possible" samples) TDE that meets the strictest post-starburst criterion, ASASSN-14li.

\begin{table}[]
    \centering
    \begin{tabular}{l l l l}
    \hline
Overenhancement & Galaxy Sample$^a$ & TDE Sample$^b$ & Source \\
\hline
33$^{+7}_{-11}\times$ & QBS & H/He broad line & [1]\\
190$^{+115}_{-100}\times$ & PSB  & H/He broad line & [1]\\
34$^{+24}_{-14}\times$ & QBS & H/He broad line & [2]\\
110$^{+80}_{-50}\times$ & PSB  & H/He broad line & [2]\\
18$^{+13}_{-9}\times$ & QBS & X-ray, likely, possible & [2]\\
18$^{+22}_{-18}\times$ & PSB  & X-ray, likely, possible & [2]\\
23$^{+21}_{-13}\times$ & QBS & X-ray, likely  & [2]\\
29$^{+41}_{-29}\times$ & PSB & X-ray, likely  & [2]\\
17$^{+12}_{-8}\times$ & QBS & Optical & [2]\\
50$^{+38}_{-29}\times$ & PSB & Optical  & [2]\\
18$^{+8}_{-7}\times$ & QBS & X-ray, likely, possible, optical  & [2]\\
35$^{+21}_{-17}\times$ & PSB &  X-ray, likely, possible, optical & [2]\\
20-80$\times$ & QBS/PSB & X-ray, likely, possible, optical & [3] \\
40-120$\times$ & QBS/PSB & X-ray, H/He broad line & [3] \\
\hline
\end{tabular}
\caption{Summary of TDE rate overenhancement found in various samples of galaxies and TDE classifications. $^a$ We note that the definitions of Quiescent Balmer-Strong (QBS) and Post-Starburst (PSB) vary slightly between \citet{French2016} and \citet{Graur2018}, and in this review we present the overenhancement as a function of the Balmer strength (see Fig \ref{fig:enhancement}). $^b$ Similarly, the TDEs used in each classification vary. For the TDEs included in the two calculations for this review, see Table \ref{tab:tde_info}. [1] \citet{French2016} [2] \citet{Graur2018} [3] This review.}
\label{tab:overenhancement}
\end{table}

We present classifications for the TDE hosts discussed in this review in Table \ref{tab:tde_sfh}, using the criteria described above. 5/41 (12\%) host galaxies are post-starburst galaxies and 13/41 (32\%) are either quiescent Balmer-strong or post-starburst. Of the 4 X-ray TDEs, 3 (75\%) are quiescent Balmer-strong and 1 (25\%) is post-starburst. Of the 15 broad H/He line TDEs, 9 (60\%) are quiescent Balmer-strong and 5 (33\%) are post-starburst.

\begin{figure*}
\includegraphics[width=0.5\textwidth]{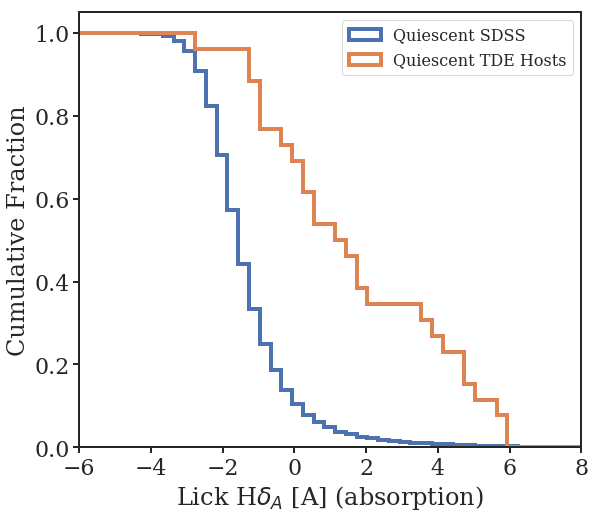}
\includegraphics[width=0.5\textwidth]{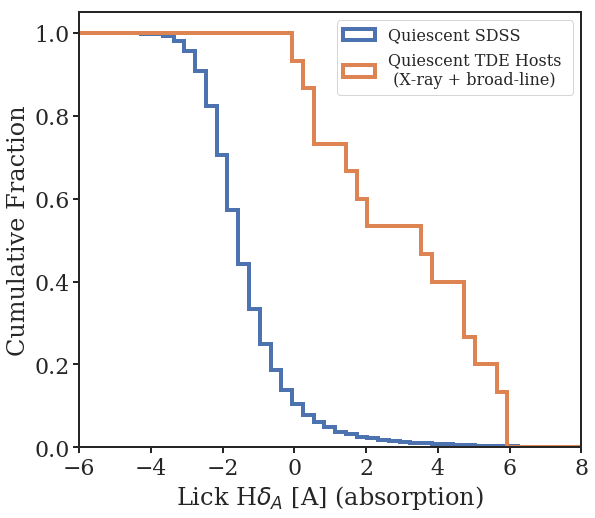}
\caption{Cumulative distribution of quiescent SDSS galaxies and quiescent TDE hosts with stronger Balmer absorption than the limit of the x-axis. For both the full set of TDE hosts considered here (left) as well as the subsets of the broad H/He line TDEs and the X-ray TDEs with the strongest post-starburst enhancement (right), there is a significant difference in the distributions between the quiescent SDSS galaxies and TDE hosts.}
\label{fig:cumulative_hd}
\end{figure*}

\begin{figure*}
\includegraphics[width=0.5\textwidth]{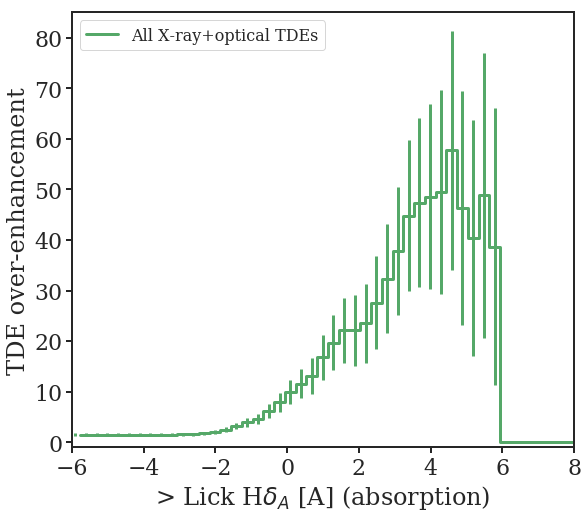}
\includegraphics[width=0.5\textwidth]{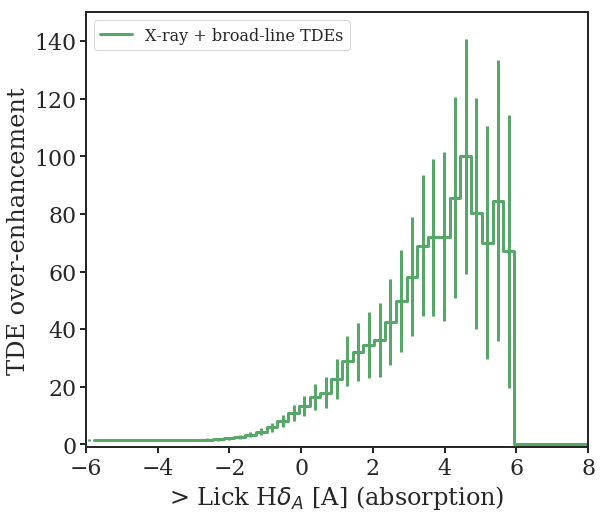}
\caption{Effect of the criteria for post-starburst or quiescent Balmer-strong on the TDE enhancement rate over normal quiescent and star-forming galaxies. For the full set of TDE hosts (left), the enhancement rate ranges for 20-80$\times$ that of normal quiescent and star-forming galaxies. For the broad H/He line TDEs and the X-ray TDEs (right), the enhancement rate ranges for 40-120$\times$ that of normal quiescent and star-forming galaxies. We note again that these TDE classifications are tentative and subject to a number of observational biases. More observations of the host galaxies for a large sample of well-characterized TDEs are needed to overcome these uncertainties and the small-number statistics limiting our precision here.}
\label{fig:enhancement}
\end{figure*}

To account for the dependence of the TDE rate enhancement on the definition of ``post-starburst" or ``quiescent Balmer-strong" we plot in Figure \ref{fig:cumulative_hd} the cumulative distribution of quiescent SDSS galaxies and quiescent TDE hosts with stronger Balmer absorption than the value on the x-axis. For both the full set of TDE hosts considered here as well as the subsets of the broad H/He line TDEs and the X-ray TDEs with the strongest post-starburst enhancement, there is a significant difference in the distributions between the quiescent SDSS galaxies and TDE hosts. We also demonstrate the effect of the criteria for post-starburst or quiescent Balmer-strong on the TDE enhancement rate over normal quiescent galaxies in Figure \ref{fig:enhancement}. For the full set of TDE hosts, the enhancement rate ranges between 20-80$\times$ that of normal quiescent galaxies. For the broad H/He line TDEs and the X-ray TDEs, the enhancement rate ranges from 40-120$\times$ that of normal quiescent and star-forming galaxies. We note again that these TDE classifications are tentative and subject to a number of observational biases. More observations of the host galaxies for a large sample of well-characterized TDEs are needed to overcome these uncertainties and the small-number statistics limiting our precision here.

\citet{Law-Smith2017} consider the post-starburst and quiescent Balmer-strong galaxy over-enhancement rates in a somewhat different subset of the \citet{Auchettl2017a} X-ray and optical TDEs, but control the galaxy parent sample on different properties to test whether selection effects drive the observed rate enhancements. Controlling on bulge mass (and thus likely black hole mass), redshift, surface brightness, \Sersic index, or bulge to total light ratio can affect the relative post-starburst or quiescent Balmer-strong galaxy rate by $\le 2\times$, within the error on the rates due to small number statistics.

Controlling on bulge colour has the largest possible effect on the observed post-starburst or quiescent Balmer-strong rates, decreasing the observed rate enhancements by factors of $\sim4$ for the post-starburst galaxies and $\sim2.5$ for the quiescent Balmer-strong galaxies, depending on what other factors are controlled for. This may be due to selection effects against finding TDEs in dustier and thus redder galaxies, or by the unique stellar populations in post-starburst galaxies, which cause them to lie in the optical green valley \citep{Wong2012}. If we control on properties which are strongly correlated with post-starburst and quiescent Balmer-strong galaxies, such as green optical colors or high central concentrations, we expect that the residual enhancement of the TDE rate in such hosts would be diminished by construction. The TDE enhancement rates in such hosts are thus consistent (given the small number statistics) between the studies of \citet{French2016}, \citet{Law-Smith2017} and \citet{Graur2018}.

While the Balmer absorption is a proxy for the recent SFH, detailed stellar population fitting can be used to determine the nature of the recent SFH using the information from the full galaxy SED. \citet{French2017} fit stellar population models to the UV/optical host galaxy photometry and optical Lick indices to determine the time elapsed since the recent starburst, the fraction of mass produced in the starburst, and the duration of the recent starburst \citep{French2018}, for a sample of host galaxies of broad H/He line TDEs. While the lower Balmer absorption ``quiescent Balmer-strong" galaxies could have had weaker H$\delta$ absorption due to longer-duration bursts, older bursts, or weaker bursts, stellar population modelling by \citet{French2017} determined that this effect is driven by the fact that quiescent Balmer-strong TDE hosts had weaker starbursts than most post-starburst galaxies. While most post-starburst galaxies have formed $\sim3-50$\% of their stellar mass in their recent starbursts, the quiescent Balmer-strong (and non-post-starburst) hosts had burst mass fractions of $0.5-2$\%. 
The post-starburst and quiescent Balmer-strong TDE hosts have ages ranging from 60 Myr to 1 Gyr since the recent starbursts ended. This large range in age suggests the enhanced TDE rate is not limited to a specific time in their host's post-starburst evolution. When compared to the total sample of post-starburst and quiescent Balmer-strong galaxies, \citet{French2017} observed a statistically insignificant dearth of TDEs in older ($>600$ Myr) post-starburst galaxies. The evolution of the TDE rate enhancement after a starburst implied by this early small sample is consistent with several models for explaining the enhanced TDE rate during this phase \citep[discussed further in \S\ref{sec:rates}]{Stone2018}.

TDEs may also be over-represented in starbursting host galaxies. However, the extreme dust extinction present in the nucleus of starburst galaxies as well as the co-existence of AGN activity make detecting such TDEs difficult. Either a lucky dust-free sightline or transient detections in the NIR are required to find TDEs in starburst galaxies. Both such scenarios have been observed. \citet{Tadhunter2017} observed a light-curve over 10 years and a serendipitous appearance of broad He lines similar to those observed in other TDEs in a starburst galaxy\footnote{This event is included in our catalogue (F01004), although its classification as a TDE is controversial. \citet{Trakhtenbrot2019} argue it may not be a true TDE, although the space for observed TDE features may be broader than expected \citep{Leloudas2019}.}. \citet{Mattila2018} observed a jet launched from the nucleus of the starburst galaxy Arp 299 believed to be caused by a TDE, with a transient discovered in NIR AO imaging. Arp 299 has a stellar population consistent with evolving to a typical post-starburst galaxy, with an starburst age of 70-260 Myr since the starburst began and 9-29\% of the total stellar mass formed in the on-going starburst \citep{Pereira-Santaella2015}. 

\citet{Tadhunter2017} estimate the TDE rate to be enhanced in such galaxies by 1000-10,000$\times$, to one per century or even one per decade per galaxy. From the observations thus far, it is unclear whether the TDE rate is enhanced during both the starburst and post-starburst phases, with selection effects biasing against observing TDEs in starburst galaxies, or if the TDE rate enhancement peak lags in time after the starburst. Upcoming infrared and radio surveys for TDEs, as well as concerted efforts to disambiguate TDEs from AGN will be necessary to resolve this question.

\subsection{Concentration and Morphology of Stellar Light; Stellar Kinematics}
\label{sec:conc}

\begin{table*}
\centering
\begin{tabular}{l c c c c c}
\hline
Name & \Sersic $n^a$ & M$_{\rm BH}$ $^b$ & A$^c$ & log $(\Sigma_{M_\star}) ^d$	& $\sigma_v ^e$ (km/s) \\
\hline
\input{tde_table_23.txt}
\hline
\end{tabular}
\caption{Table of properties of known TDE host galaxies discussed in \S\ref{sec:conc}. $^a$ \Sersic indices from \citet{Mendel2014} fits. $^b$ Black hole masses from \citet{Law-Smith2017}. $^c$ Residual asymmetry measures from \citet{Simard2011}. $^d$ Surface stellar mass density, computed as ${{\rm{\Sigma }}}_{{M}_{\star }}=\mathrm{log}[({M}_{\star }/{r}_{50}^{2})/({M}_{\odot }/{{\rm{kpc}}}^{2})]$ by \citet{Graur2018}. $^e$ Stellar velocity dispersions from SDSS Portsmouth pipeline or \citet{Wevers2017}. $\dagger$ Broad line TDEs (see Table \ref{tab:tde_info}). $^*$ Coronal line TDEs (see Table \ref{tab:tde_info}).
}
\label{tab:conc}
\end{table*}

The TDE rate is expected to depend on various physical properties of the SMBH and the stellar population in its vicinity, including the mass of the SMBH, the density of stars within its loss cone, and their velocity dispersion. Unfortunately, it is exceedingly hard to observationally probe the parsec-scale region of influence, except for the most nearby SMBHs. However, some global host-galaxy properties, on kpc scales, are known to be correlated with local properties in galactic nuclei. The most fundamental of these is the $M$--$\sigma$ relation, which relates the mass of the SMBH, $M$, to the host galaxy stellar velocity dispersion, $\sigma$ (e.g., \citealt{Kormendy1995,Magorrian1998,2000ApJ...539L..13G,Tremaine2002,McConnell2013}). Moreover, these central stellar velocity dispersions have been shown to be correlated over galactic scales \citep{Cappellari2006}.

\begin{figure*}
\includegraphics[width=0.5\textwidth]{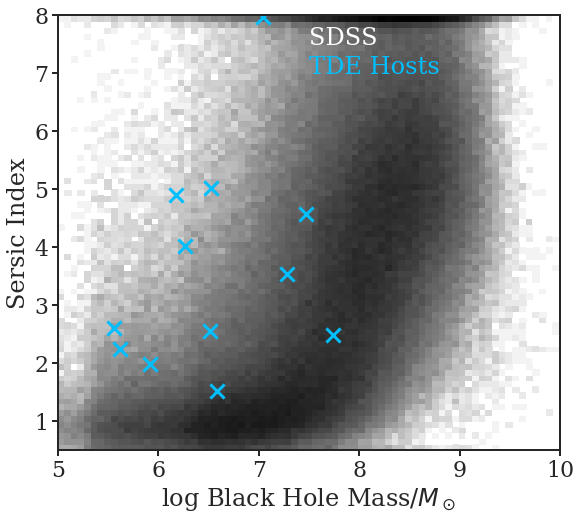}
\includegraphics[width=0.5\textwidth]{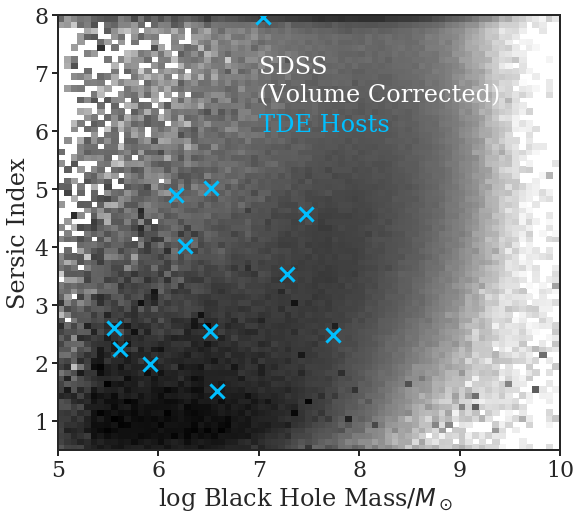}
\includegraphics[width=0.5\textwidth]{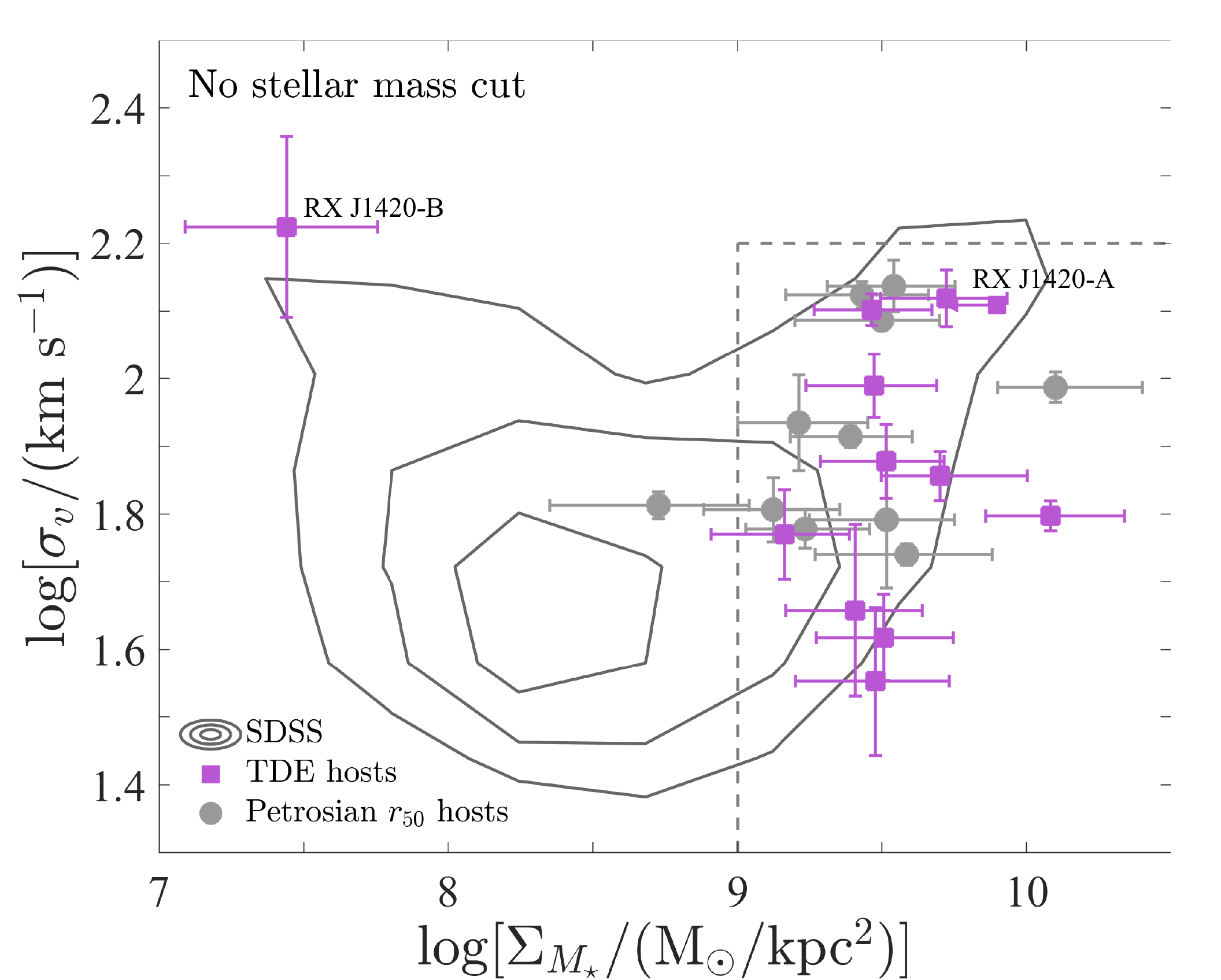}
\includegraphics[width=0.5\textwidth]{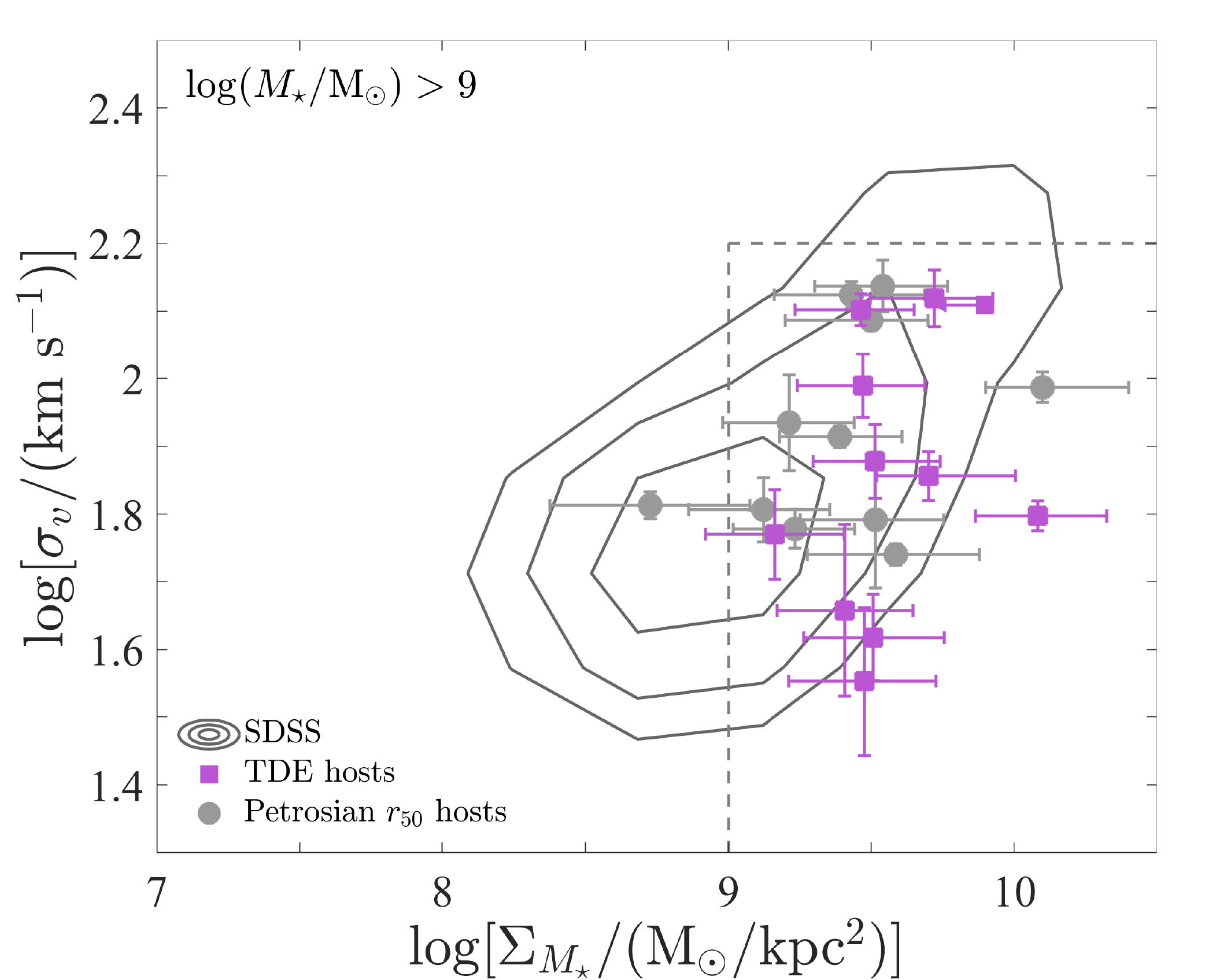}
\caption{
{\bf Top Left:} Black hole masses inferred from the $M-\sigma$ relation vs. \Sersic indices from \citet{Simard2011} for the SDSS main galaxy sample and the SDSS TDE hosts, adapted from \citep{Law-Smith2017}. The TDE hosts have high \Sersic indices for their black hole masses. {\bf Top Right:} Adapted from \citet{Law-Smith2017} with volume corrections from \citet{Mendel2014} as done in the bottom left plot. Despite the large number of low black hole mass galaxies inferred from the volume correction, the TDE hosts still lie at high \Sersic indices for their black hole masses, compared with the rest of the galaxy sample. {\bf Bottom Left:} Updated from \citet{Graur2018}, TDE host galaxies (symbols) significantly prefer galaxies with high surface stellar mass densities, relative to a volume-limited SDSS sample (contours). New measurements for NGC 5905 and SDSS J1201 have been added to these plots. Contours represent the 25th, 50th, and 84th percentiles of the volume-weighted background galaxy distribution. Host galaxies marked by purple squares have galaxy properties measured by the same pipeline as the SDSS control sample. Host galaxies marked by grey circles differ from the SDSS control sample because their stellar surface densities are measured using Petrosian radii rather than \Sersic half-light radii, and some have velocity dispersions measured by \citet{Wevers2017} instead of the SDSS pipeline.  The TDE host galaxies are significantly denser than the control sample (although this trend depends on whether the galaxies are star-forming or not -- see text). 
{\bf Bottom Right:} (Reproduced by permission of the AAS.) From \citet{Graur2018}, same as the previous panel, but with a stellar mass cut to include only galaxies with similar stellar masses as the TDE host galaxies. Contours again represent the 25th, 50th, and 84th percentiles of the volume-weighted background galaxy distribution. As before, the TDE host galaxies have a significantly higher distribution of stellar surface mass densities, even after removing the sample of low mass, low surface density galaxies magnified by the volume correction. While the axes in these plots are not strictly analogous to those in the top plots, both analyses demonstrate that the TDE hosts have higher stellar concentrations than expected given their velocity dispersion or inferred black hole mass. Data from this Figure can be found in Table \ref{tab:conc}.
}
\label{fig:concentration_summary}
\end{figure*}

\citet{Graur2018} compared a sample of 11 TDE host galaxies with surface stellar mass densities and velocity dispersions computed from galaxy properties measured by the Sloan Digital Sky Survey (SDSS; \citealt{York2000,Kauffmann2003,Brinchmann2004}) to a volume-limited sample of SDSS galaxies with galaxy properties measured by the same pipeline. Their TDE host galaxies had surface stellar mass densities in the range $\Sigma_{M_\star}=10^{9-10}~M_\odot~{\rm kpc}^2$. The star-forming TDE hosts were significantly denser than the star-forming control sample. This effect was not significant for quiescent galaxies, which already tend to have high surface stellar mass densities. \citet{Graur2018} also measured surface stellar mass densities for a similar sample of 9 TDE host galaxies with velocity dispersions measured by \citet{Wevers2017}, and found that they too had values in the range $10^{9-10}~M_\odot~{\rm kpc}^2$ with one exception: PS1-10jh, which had a surface stellar mass density of ${\rm log}(\Sigma_{M_\star}/M_\odot~{\rm kpc}^2)=8.7^{+0.3}_{-0.4}$. Both of these samples, in purple and gray markers, respectively, are shown in Figure~\ref{fig:concentration_summary}. Because the volume-corrected quiescent galaxies have a different stellar mass distribution than the star-forming galaxy sample, we also compare the $\Sigma_{M_\star}$ and velocity dispersion for a comparison sample cut in stellar mass to be M$_\star > 10^9$ M$_\odot$ in order to match the TDE host galaxies. Even after removing the low mass galaxies, the TDE hosts still have higher stellar surface densities than the volume-corrected comparison sample. 

While the preference of TDE hosts for galaxies with high surface stellar mass densities was statistically significant (at least for star-forming galaxies), there was no significant dependence on the galaxy central stellar velocity dispersion. Only the quiescent host galaxies showed a hint that their velocity dispersions might be lower than those of the quiescent galaxies in the control sample. It remains to be seen whether this effect proves to be significant in a larger sample.

\citet{Law-Smith2017} also compared kpc-scale indicators of stellar mass concentration for a sample of 10 TDE hosts with data from the \citet{Simard2011} and \citet{Mendel2014} SDSS catalogues, comparing the \Sersic indices of the TDE host galaxies and SDSS comparison galaxies to the black hole masses inferred from the $M-\sigma$ relation. The TDE host galaxies have \Sersic indices in the top 10--15\% of the comparison sample in bins of black hole mass, indicating the TDE host galaxies have more concentrated stellar populations. In this review, we add data for two new TDEs (ASASSN-18zj and AT2018dyk) with archival SDSS information. We also perform a volume correction for the SDSS comparison sample using the volume calculations of \citet{Mendel2014} in order to compare this analysis to that of \citet{Graur2018}. The updated \Sersic index--black hole mass plots are shown in Figure \ref{fig:concentration_summary}. The volume correction accounts for the larger number of galaxies with low black hole mass, but the same trend of TDE host galaxies having higher \Sersic indices for their black hole masses is seen. We find that 50\% of the TDE host galaxies have \Sersic indices in the top 20\% of the volume-corrected SDSS galaxies with M$_{\rm BH}$ $10^5-10^6$ M$_\odot$, top 10\% of the volume-corrected SDSS galaxies with M$_{\rm BH}$ $10^6-10^7$ M$_\odot$, and top 30\% of the volume-corrected SDSS galaxies with M$_{\rm BH}$ $10^7-10^8$ M$_\odot$. If we only compare to the volume-corrected SDSS galaxies with quiescent levels of star formation (SFR$<1$ M$_\odot$ yr$^{-1}$), we find the same result for M$_{\rm BH}$ $10^5-10^7$ M$_\odot$, but no significant enhancement of TDE hosts in higher \Sersic index galaxies with black hole masses M$_{\rm BH}$ $10^7-10^8$ M$_\odot$.

A similar trend is also seen if the bulge to total light ratio is used as a proxy for stellar concentration instead of the \Sersic index \citep{Law-Smith2017}. However, the \Sersic index measurements from \citet{Simard2011} have lower errors and thus allow finer binning and a more detailed comparison for our analyses.

The analyses by \citet{Law-Smith2017} and \citet{Graur2018} have established that TDE hosts are more concentrated on galaxy-wide (kpc) scales. We discuss possible mechanisms for the stellar concentration affecting the TDE rate, and the interplay between this effect and the trend with star formation history in Section \ref{sec:rate_conc}, as post-starburst and quiescent Balmer-strong galaxies are also known to have high central concentrations of stellar light.

We also consider here the quantitative morphologies of the TDE host galaxies in asymmetry - concentration space. Spiral galaxies and elliptical galaxies separate in this space with elliptical galaxies having higher concentrations and spiral galaxies having higher asymmetries \citep{Abraham1996}. Mergers can be further identified, with higher asymmetries than individual galaxies \citep{Conselice2003}. Post-starburst galaxies often show signs of recent mergers, but at several hundred Myr past coalescence, their asymmetries as measured using HST imaging have lessened to be between those of elliptical and spiral galaxies \citep[Figure 5 reproduced in this review]{Yang2008}.

\begin{figure*}[h!]
\includegraphics[width=0.5\textwidth]{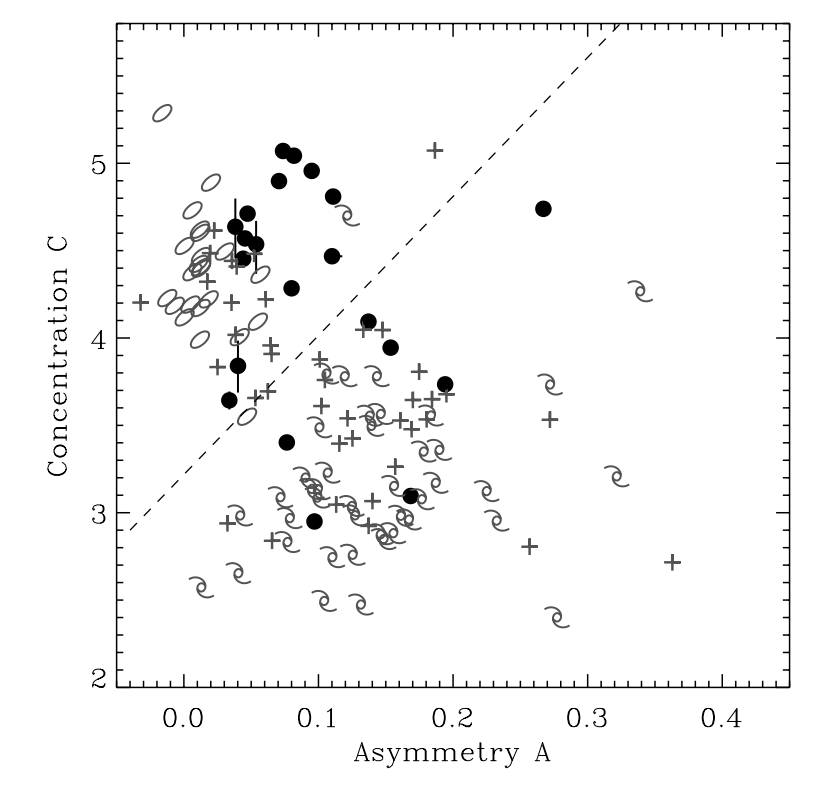}
\includegraphics[width=0.5\textwidth]{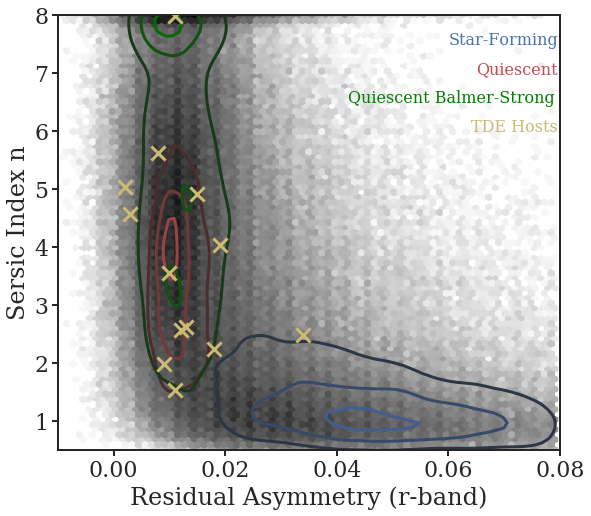}
\caption{Left: (Reproduced by permission of the AAS.) From \citet{Yang2008}, showing the concentration and asymmetry trends for star-forming (spiral symbols), quiescent (elliptical symbols), intermediate spirals (plus signs), and post-starburst galaxies (filled circles) from HST imaging. The post-starburst sample has high concentrations of stellar light, comparable to or higher than the quiescent galaxies, and asymmetries stronger than the quiescent galaxies, but less pronounced than the star-forming galaxies. Right: Adapted from \citet{Law-Smith2017}, we plot the \Sersic index vs. the $r$-band residual asymmetry. In contrast to the left-hand panel, the residual asymmetries are calculated after the best-fit smooth galaxy model has already been subtracted. The residual asymmetries are calculated by \citet{Simard2002,Simard2011} and are denoted RA1\_2 in the GIM2D output tables. Contours are plotted for the subsamples of star-forming (blue), quiescent (red), and quiescent Balmer-strong galaxies (green). TDE hosts are plotted as yellow crosses. The star-forming galaxies have high asymmetries and low \Sersic indices, whereas the quiescent galaxies have \Sersic indices of $\sim3-5$ and low asymmetries. The quiescent Balmer-strong galaxies show high \Sersic indices, with a tail extending down to the quiescent galaxies. The TDE hosts follow the quiescent galaxy distribution, with one outlier (SDSSJ0952) at high \Sersic index ($n\sim8$). Higher resolution imaging is required to determine whether the TDE host galaxies show the trend towards intermediate asymmetries indicative of recent mergers, as seen in the HST imaging of post-starburst galaxies. Data from this panel can be found in Table \ref{tab:conc}.
}
\label{fig:c_a}
\end{figure*}

\citet{Law-Smith2017} have compiled a sample of morphological indicators for the TDE host galaxies as well as the SDSS main spectroscopic sample, using the catalogues of \citet{Simard2011}. We compare the concentration to the residual asymmetries \citep{Simard2002} for the SDSS galaxies and TDE host galaxies in Figure \ref{fig:c_a}, using the \Sersic index as a proxy for concentration. We separate out star-forming, quiescent, and quiescent Balmer-strong galaxies to identify trends in this space. The star-forming galaxies have high asymmetries and low \Sersic indices, while the quiescent galaxies have \Sersic indices of $\sim3-5$ and low asymmetries. The quiescent Balmer-strong galaxies show high \Sersic indices\footnote{See above for a more thorough discussion of the \Sersic indices of TDE hosts, accounting for trends in stellar mass and black hole mass.}, with a tail extending down to the quiescent galaxies, and low asymmetries. The TDE hosts are distributed like the quiescent galaxies, with one source (SDSSJ0952) having a high \Sersic index of $n\sim8$. The shift towards higher asymmetry for the post-starburst or quiescent Balmer-strong galaxies is not observed in the residual asymmetries from the SDSS imaging as it was in the total light asymmetries from the HST imaging; this is likely due to the greater sensitivity of the HST data to low surface brightness tidal features. \citet{Yang2008} found that many of the tidal features observed with HST imaging would not be observable with ground-based imaging. Thus, the lack of high asymmetries in the TDE host sample does not rule out a recent merger, even a recent major merger. Higher resolution imaging and a variety of new measures of galaxy asymmetry \citep[e.g.,][]{Pawlik2015} will be required to determine whether the TDE host galaxies have the trend towards intermediate asymmetries indicative of recent mergers, as seen in the HST imaging of post-starburst galaxies.

\begin{landscape}
\begin{table*}
\centering
\begin{tabular}{l c c c c c c l l}
\hline
Name & H$\alpha$ & H$\beta$ & [NII]6584 & [SII]6717+6731 & [OIII]5007 & BPT Class$^a$ & W1-W2$^b$ & SDSS Notes$^c$ \\
\hline
\input{tde_table_bpt.txt}
\hline
\end{tabular}
\caption{Table of properties of known TDE host galaxies discussed in \S\ref{sec:agn}. Fluxes are in units of $10^{-17}$ ergs/s/cm$^2$. $^a$ Classification from [NII]-BPT diagram. $^b$ WISE 3.4$\mu$m - 4.6$\mu$m colours. We \textbf{bold} those values that meet the \citet{Stern2012} WISE AGN criteria. $^c$ SDSS {\tt targettype} and {\tt subclass} notes indicating the presence of broadline AGN or QSO like spectra. $^d$ While the host galaxy of PS16dtm is classed as SF on the [NII]-BPT diagram, it is a Seyfert I \citep{Blanchard2017}. $\dagger$ Broad line TDEs (see Table \ref{tab:tde_info}). $^*$ Coronal line TDEs (see Table \ref{tab:tde_info}).}
\label{tab:agn}
\end{table*}
\end{landscape}

\subsection{AGN Activity}
\label{sec:agn}

\begin{figure}
\includegraphics[width=0.5\textwidth]{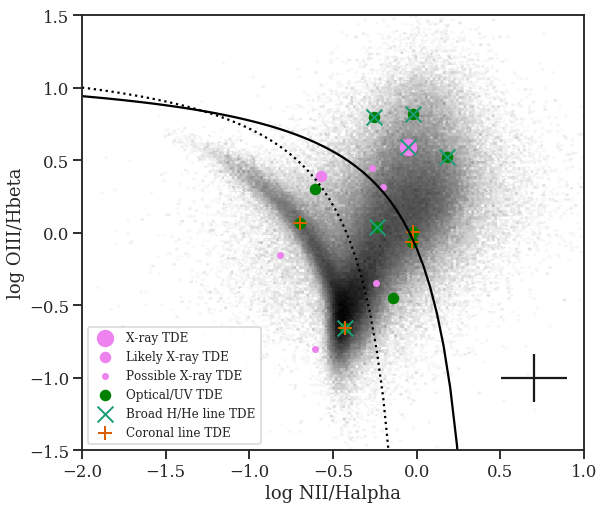}
\includegraphics[width=0.5\textwidth]{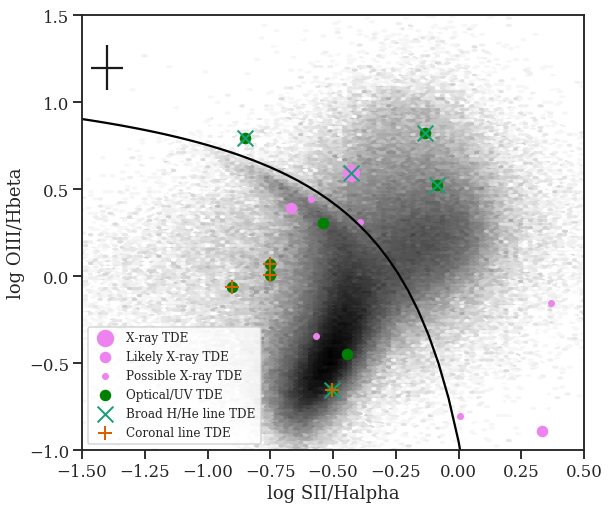}
\caption{BPT \citep{Baldwin1981} diagrams of emission line ratios indicative of ionisation from AGN or star formation in the TDE host galaxies and SDSS main galaxy spectroscopic sample. Galaxies to the lower left of the dotted and solid lines have ionisation dominated by star formation. Galaxies to the upper right of the dotted lines have ionisation dominated by AGN, although those with relatively weak emission lines may instead have ionisation from shocks or evolved stars \citep[e.g.,][]{Rich2015, Yan2012}. The dotted line is the observed star formation - AGN separation from \citet{Kauffmann2003b} and the solid lines are the theoretical maximum starburst lines from \citet{Kewley2001}. Characteristic error bars are shown in each panel. The TDE host galaxies occupy a range of star-formation dominated, AGN-dominated, and ambiguous hosts. 
}
\label{fig:bpt}
\end{figure}

We consider here the possibility that some TDEs may occur in an environment with a pre-existing accretion disk. We have compiled a BPT \citep{Baldwin1981} diagram in Figure \ref{fig:bpt} showing the TDE host galaxies as well as galaxies from the SDSS main spectroscopic survey. We include TDE host galaxies with SDSS spectra as described above, as well as galaxies with emission line ratios measured by \citet{French2017} and \citet{Wevers2019}. These emission line fluxes are shown in Table \ref{tab:agn}. We classify galaxies into star-forming, composite, and AGN Seyfert II or LINER based on the classifications of \citet{Kewley2001} and \citet{Kauffmann2003b}. These classifications are subject to a number of caveats, and represent the likely dominant ionisation source in the aperture probed by the spectrum. However, many galaxies in the AGN region of the BPT diagram, especially those with relatively weak emission lines, may instead have ionisation consistent with an origin from shocks or evolved stars \citep[e.g.,][]{Rich2015, Yan2012}. ``Composite" galaxies lie in between the star-forming and AGN regions and could be a mix of star-formation and other ionisation sources. The TDE host galaxies occupy a range of star-formation dominated, AGN-dominated, and ambiguous ionisation source galaxies.

Another way of identifying AGN, especially those obscured by dust, is to look for signatures of hot dust from the WISE 3.4--4.6$\mu$m colours. \citet{Stern2012} identify a WISE colour cut of WISE 3.4--4.6$>0.8$ Vega mag to indicate the presence of an AGN. We present these WISE colours in Table \ref{tab:agn}, identifying four TDE hosts which meet this criterion: the hosts of F01004, SDSS J0952, SDSS J1342, and SDSS J1350. The first host galaxy is currently experiencing a starburst \citep{Tadhunter2017}. The latter three galaxies all hosted TDEs with observed coronal line emission.

The BPT analysis described above selects narrow-line AGN (Seyfert II galaxies), or obscured AGN, as does the infrared selection. When the broad-line regions of AGN are visible, this provides another way to identify them from their optical spectra. In Table \ref{tab:agn}, we also list notes from the SDSS to indicate broad-line or QSO emission. Five TDE host galaxies have such notes, with varying overlap with those galaxies classified as AGN from the BPT or WISE colour analyses. AGN selection using any of these methods is neither pure nor complete. One may note that the host galaxy of PS16dtm, a Narrow-line Seyfert I galaxy \citep{Blanchard2017}, is not selected as an AGN using a BPT or WISE colour analysis, and requires further analysis of the optical spectrum to identify the broad components of the Balmer lines. The connection between TDEs and Narrow line Seyfert I galaxies requires further study \citep{Wevers2019b}. Such galaxies make up only $\sim15$\% of all Seyfert I galaxies \citep{Williams2002} and have been observed to have optical flares similar to TDEs \citep{Kankare2017}. Some of these events may even belong to a different class of transient \citep{Frederick2019}.

Caution should be taken in interpreting these results given the selection effects against identifying TDEs in AGN host galaxies. We discuss the role of AGN in either enhancing the TDE rate or as the source of selection effects against identifying TDEs in such host galaxies in \S\ref{sec:circumnucleargas}. TDEs will be more difficult to identify in AGN due to selection against AGN flares and higher levels of dust obscuration. These effects may also bias the types of AGN TDEs are found in. Further study will be needed to fully understand these effects.

Furthermore, we note that spectra taken after the TDE may be contaminated by residual TDE emission, depending on how long the emission persists for. \citet{Brown2016b} found that narrow H$\alpha$ emission can persist for a year after the TDE, but after several years, \citet{Wevers2019} find no residual narrow line emission. \citet{French2017} noted a tentative offset in H$\alpha$ equivalent width and [NII]-6584/H$\alpha$ emission ratio between the few events with spectroscopy before vs. after the TDE. \citet{French2017} found the host galaxies with spectroscopy from after the TDE to have higher H$\alpha$ equivalent widths and [NII]-6584/H$\alpha$ emission ratios than the host galaxies with spectroscopy from before the TDE, although this analysis was limited by the small number of events and the lack of events with spectra from both before and well-after the TDEs. In the sample considered in this review, we note the host galaxies with spectra taken before the TDE contain more Seyferts and the host galaxies with spectra taken after the TDE contain more star-forming and composite classifications. However, this comparison is still limited by small number statistics and selection effects between the various TDE detection methods used. A systematically collected set of follow-up spectra will be needed to better understand the presence of narrow line emission in the decade after a TDE.

\begin{figure}
\includegraphics[width=1\textwidth]{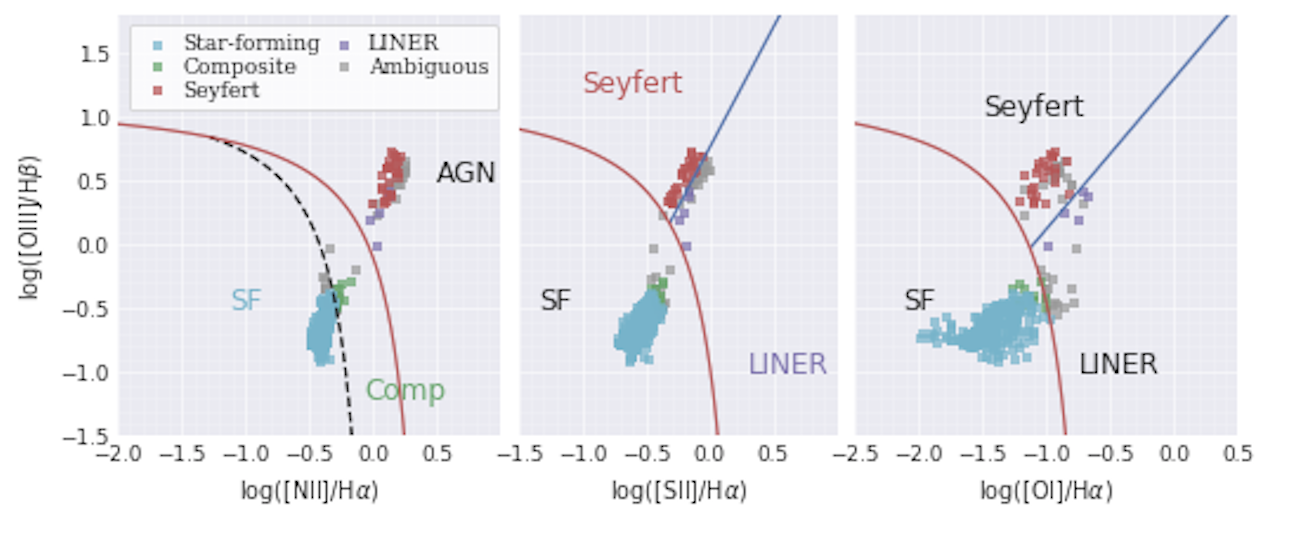}
\includegraphics[width=0.32\textwidth]{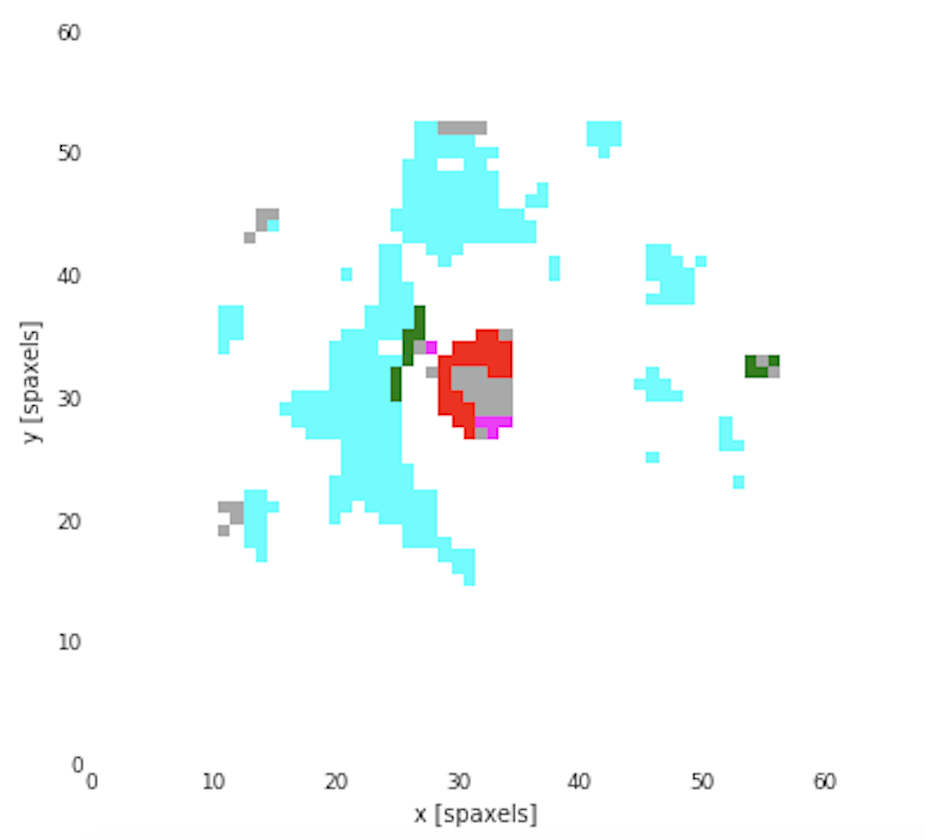}
\includegraphics[width=0.32\textwidth]{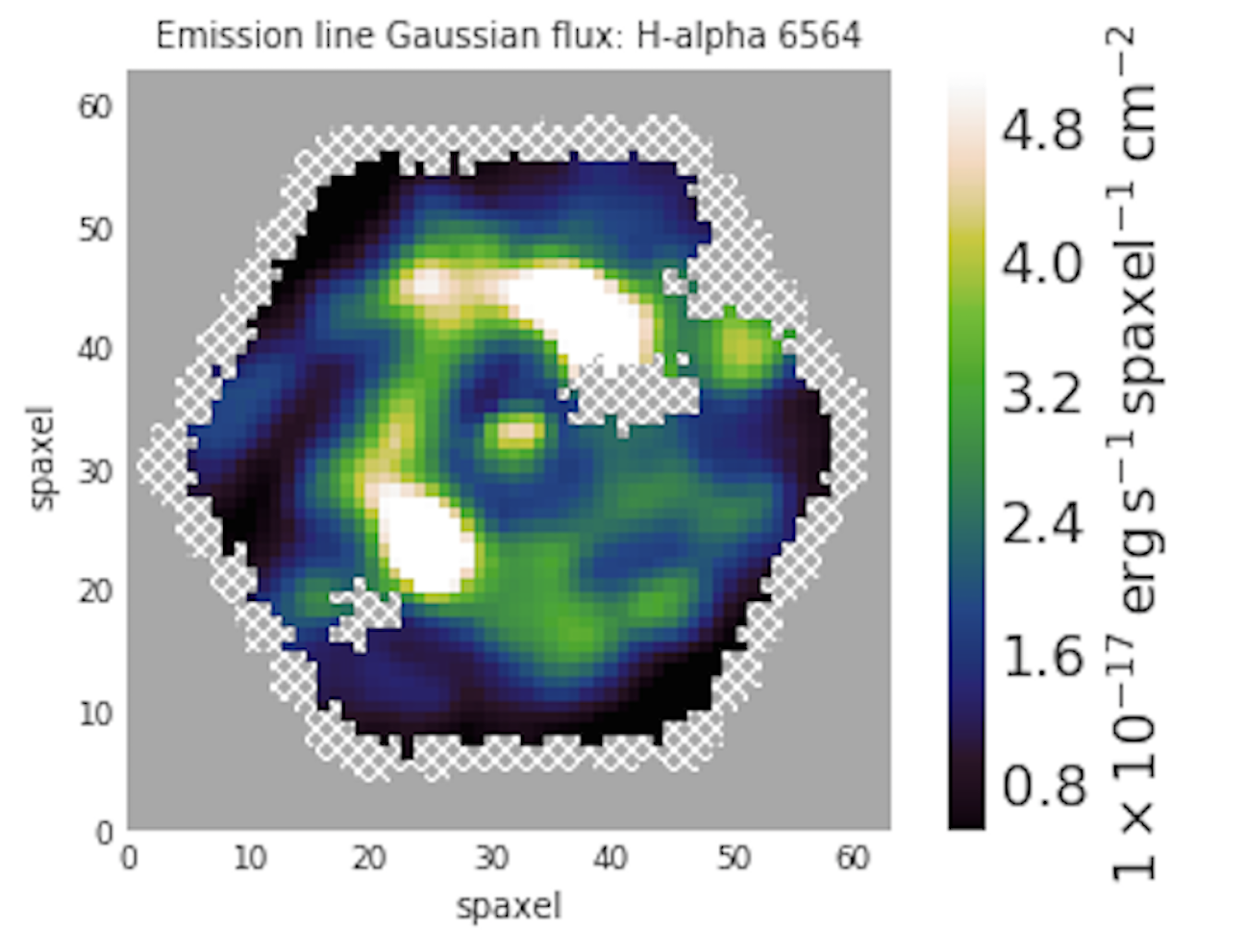}
\includegraphics[width=0.32\textwidth]{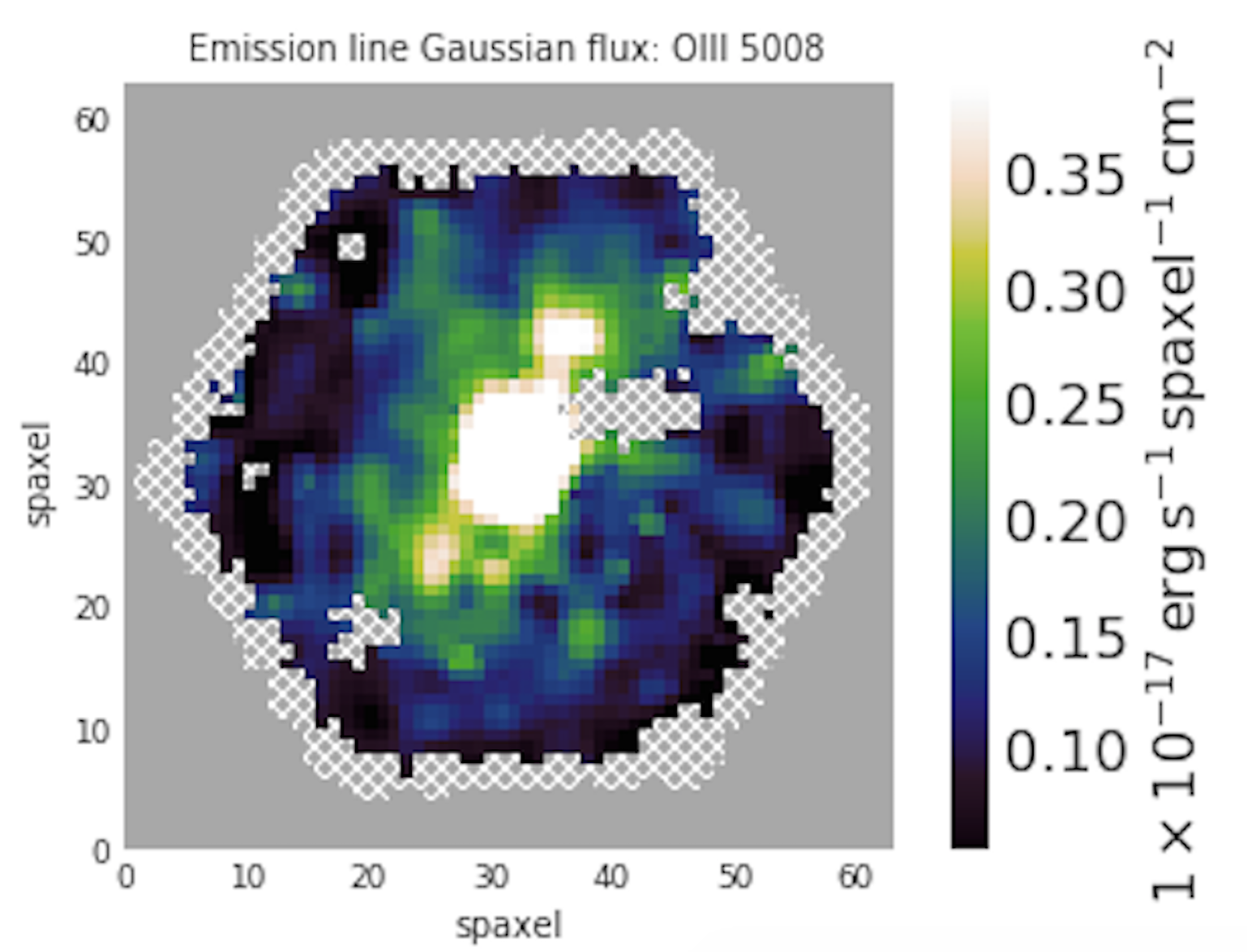}
\caption{Spatially resolved BPT diagram, H$\alpha$ map, and [OIII]5007 map for the host galaxy of TDE AT2018dyk, made using Marvin \citep{marvin}. Points are colored blue, green, red, dark grey, or light grey based on their classification as star-forming, composite, Seyfert, LINER, or ambiguous. This host galaxy is in the AGN regions of the BPT diagrams in Figure \ref{fig:bpt}, and here we see that there is indeed a central [OIII]-bright source, and that the outskirts of the galaxy have ionisation dominated by star formation. Obtaining data like this for additional TDE hosts, especially those for which the ionisation source is ambiguous, or there is evidence for unusual AGN properties, will help further our understanding of the co-existence of TDEs and AGN.
}
\label{fig:manga}
\end{figure}

Further insight into the presence or absence of AGN in TDE host galaxies requires spatially resolved emission line maps from IFU data. We present one example here; the host galaxy of AT2018dyk was observed as part of the MaNGA survey \citep{manga}. This host galaxy is in the AGN regions of the BPT diagrams in Figure \ref{fig:bpt}, and in Figure \ref{fig:manga} we see that there is indeed a central [OIII]-bright source, and that the outskirts of the galaxy have ionisation dominated by star formation.

\begin{figure}
\begin{center}
\includegraphics[width=0.5\textwidth]{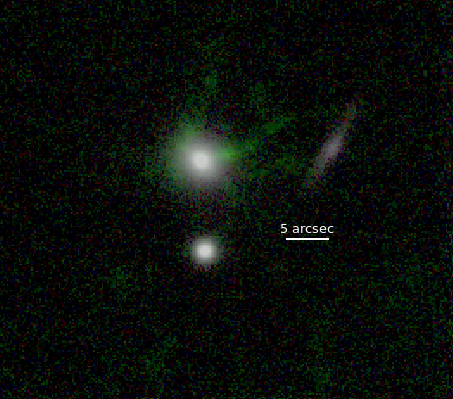}
\caption{Three-colour RGB image constructed from a MUSE datacube of the host galaxy of ASASSN-14li. The red and blue components of this RGB image are from red and blue continuum regions and green is from the [OIII]~5007 line. Extended ionized features are observed via the [OIII]~5007 line extending beyond the continuum-dominated bulk of the galaxy, and are likely from past AGN activity $10^4-10^5$ years ago \citep{Prieto2016}. The star observed below the galaxy and the edge-on galaxy to the right are not associated with the TDE host galaxy. Recent AGN activity in galaxies lacking strong current AGN activity further complicates our understanding of the co-existence of gas and stellar accretion by supermassive black holes in galaxies, raising questions of the timescale for TDE rate enhancements compared to AGN duty cycles.}
\label{fig:muse}
\end{center}
\end{figure}

The host galaxy of the TDE ASASSN-14li additionally shows evidence of past AGN activity, with large extended ionized regions visible in [OIII]~5007 and [NII]~6584 lines, seen in MUSE observations \citep[Figure \ref{fig:muse}]{Prieto2016}. Given the light travel time to these narrow-line regions, their ionisation implies strong AGN activity $10^4-10^5$ years in the past. Such extended ionized features have been seen around other galaxies in large imaging surveys \citep[``voorwerps"][]{Lintott2009, Keel2017} and around other post-starburst galaxies in narrow-band imaging \citep{Schweizer2013, Watkins2018}. These instances of recent AGN activity in galaxies lacking strong current AGN activity further complicate our understanding of the co-existence of gas and stellar accretion by supermassive black holes in galactic nuclei, raising questions regarding the timescale for TDE rate enhancements compared to AGN duty cycles. The host galaxy of ASASSN-14li furthermore has a persistent radio source discovered in the FIRST survey which may also indicate on-going low-level AGN activity \citep{Holoien2016}.

\subsection{Environment}
\label{sec:enviro}

We consider here the extragalactic environments of the TDE host galaxies, as many mechanisms which act to change the other galaxy properties considered in this section can only act in dense cluster-scale environments. Several TDEs have been found in targeted X-ray surveys of dense galaxy clusters. J1311 \citep{Maksym2010} was found in a search for TDEs in Abell 1689. While it is not included in our analysis due to the lack of a host galaxy spectrum, ``Wings" \citep{Maksym2013} was found in Abell 1795. 

In addition to possible selection biases towards finding TDEs in clusters, post-starburst galaxies are known to lie preferentially in group environments \citep{Zabludoff1996}. While groups like the Local Group are the most common galaxy environment, the groups favored by post-starburst galaxies tend to be virialized and more massive than the Local Group, with low enough velocity dispersions and high enough galaxy densities to make galaxy-galaxy mergers and tidal interactions likely \citep{Zabludoff1996}. 

We cross match the catalogue in Table \ref{tab:tde_info} with the Abell cluster catalogue \citep{Abell1989}, to check for host galaxies within 25 arcminutes of a cluster with a similar redshift (velocities within 3000 km/s of the host galaxy redshift). We find one additional TDE host galaxy, PTF09axc, to be associated with the cluster Abell 1986. 

To test for TDE host galaxies in less rich clusters and groups, we cross match our TDE host galaxy catalogue with the group and cluster catalogue of \citet{Tempel2014}. Of the 41 host galaxies, 14 are matched with the \citet{Tempel2014} catalogue. Seven are the only galaxy in their halo\footnote{ASASSN14li, RBS1032, SDSS J1342, SDSS J0748, ASASSN14ae, ASASSN18zj} , six are associated with groups of 1-6 additional galaxies\footnote{SDSS J1323, NGC 5905, SDSS J0952, PTF09ge, AT2018dyk, AT2018bsi}, and one is part of a cluster of 52 galaxies (SDSS J1350). Given the set of 14 TDE host galaxies in the \citet{Tempel2014} catalogue, and their full sample of galaxies, we find no evidence to suggest that the TDE host galaxies prefer different environments than the general galaxy sample. Using either a Kolmogorov-Smirnov or Anderson-Darling test, we cannot reject the hypothesis their environment richnesses are drawn from the same distribution.

\section{Possible Drivers for Host Galaxy Preferences}
\label{sec:discussion}

There are several possible causes for the observed TDE rate enhancement in the host galaxy types discussed above. Many of these scenarios predict TDE enhancements in both post-starburst hosts and centrally concentrated hosts, such as a central overdensity or mechanisms related to galaxy-galaxy mergers. 

\subsection{Increased Stellar Concentration / Central Overdensities}
\label{sec:rate_conc}
The TDE rate depends on the number of stars which can be scattered into center-crossing orbits, and so a high central stellar density will result in a high TDE rate. For a Nuker surface-brightness profile, the inner stellar slope $\gamma$ is found to correlate with the TDE rate as $\dot{N}_{\rm TDE} \propto \gamma^{0.705}$ in a sample of early type galaxies from \citet{Lauer2007} \citep{Stone2016b}. A high central stellar concentration may be correlated with high concentrations on larger $\sim$kpc scales. As discussed above in \S\ref{sec:conc}, TDEs are overrepresented in galaxies with high \Sersic indices for their black hole masses \citep{Law-Smith2017} and in galaxies with high stellar surface densities on scales of the half-light radius \citep{Graur2018}. An increased TDE rate due to a merger-induced stellar overdensity is seen in simulations by \citet{Pfister2019}, although at very early stages in the merger, before the coalescence of the two black holes.

If high stellar concentrations drive the TDE rate enhancement in high \Sersic index or stellar surface density galaxies, it may also explain the rate enhancement in post-starburst galaxies. Post-starburst galaxies have high \Sersic indices \citep{Yang2004, Quintero2004, Yang2008} as the recent starbursts are centrally concentrated, likely due to stars formed from gas infall in the recent merger, and the young/intermediate A stars dominate the light. Once the bright young stellar population in post-starburst galaxies fades, the bulge properties are consistent with evolving to normal early type galaxies, but the stellar concentrations on scales close to the black hole radius of influence have not been measured in samples of post-starburst galaxies. However, the nearby post-starburst galaxy NGC 3156 has HST imaging with high enough resolution to measure the central slope of the Nuker profile, and \citet{Stone2016} find the slope to be steeper than any of the early type galaxies studied previously by \citet{Stone2016b}. Given the lack of a similar TDE rate enhancement in early type galaxies, something must change in the central galaxy concentration or dynamics in the few Gyrs after the post-starburst phase. 

The evolution of a central overdensity with time was studied by \citet{Stone2018}, who model the stellar density profile as $\rho \propto r^{-\gamma}$, and determine how $\gamma$ changes with time. Given the post-burst ages of the post-starburst TDE hosts \citep{French2017}, the TDE rate enhancement and its tentative evolution with time could be explained if $\gamma \ge 2.5$. The predictions made by this model can be tested with larger samples of post-starburst TDE hosts in the LSST and perhaps even ZTF eras. 

Based on the supposition that the TDE rate should depend on the density of stars in the SMBH loss cone, along with their velocity dispersions, \citet{Graur2018} assumed those local properties would be correlated with their global, kpc-scale counterparts, and that the TDE rate would depend on the latter as $R_{\rm TDE}\propto \Sigma_{M_\star}^\alpha \times \sigma_v^\beta$, where $\Sigma_{M_\star}$ is the surface stellar mass density on the scale of the half-light radius and $\sigma_v$ is the kpc-scale velocity dispersion. By comparing their sample of TDE host galaxies with a volume-limited control sample drawn from the SDSS, \citet{Graur2018} estimated the values of the power-law indices to be $\hat{\alpha}=0.9\pm0.2$ and $\hat{\beta}=-1.0\pm0.6$ using SDSS fiber measurements of the central few kpc, and assuming these global properties correlate with the properties on the smaller scales of the stars in the SMBH loss cone. 

\citet{Wang2004} find that the TDE rate of an isothermal sphere ($\rho \propto r^{-2}$) depends on the SMBH mass, $M_\bullet$, and local velocity dispersion, $\sigma$, as $R_{\rm TDE}\propto M_\bullet^{-\alpha}\times \sigma^\eta$. The average surface stellar mass density of the stars orbiting the SMBH is $\Sigma = M_\bullet/\pi r_h^2=\sigma^4/\pi G^2 M_\bullet$, where $r_h=GM_\bullet /\sigma^2$ is the size of the star cluster \citep{Peebles1972}, and $G$ is the gravitational constant. This allows us to rewrite the TDE rate as $R_{\rm TDE}\propto \Sigma^\alpha \times \sigma^{\eta-4\alpha}$. Using the \citet{Graur2018} estimates for $\alpha$ and $\beta=\eta-4\alpha$, the values measured from the data, $\alpha=0.9\pm0.2$ and $\eta=2.6\pm1.0$ are consistent with the theoretical predictions, $\alpha=1$ and $\eta=3.5$ \citep{Wang2004}. This suggests that the TDE rate is indeed driven by the dynamical relaxation of stars into the loss cone of the SMBH.


Could both the preference of TDE hosts to be in post-starburst or quiescent Balmer-strong host galaxies and the preference for host galaxies with high central concentrations be driven by the same effect? Similar galaxy over-representations can be found in both \Sersic index--black hole mass and in H$\alpha$ emission--H$\delta$ absorption, where $\ge 60$\% of TDEs are found in $\sim2$\% of the parameter space (i.e., at high H$\delta$ absorption and low H$\delta$ emission, or at high \Sersic index and low black hole mass)\footnote{Although the details of this will depend on the TDE samples and comparison samples used, see previous discussions in \S\ref{sec:sfh} and \S\ref{sec:conc}.}. Of the five TDE host galaxies considered by \citet{Law-Smith2017} with high \Sersic indices and low black hole masses, three are post-starburst or quiescent Balmer-strong, and two (PTF09ge and SDSSJ123) are not. Of the seven quiescent Balmer-strong galaxies considered by \citet{French2016}, two (ASASSN-14li and ASASSN-14ae) also meet the high \Sersic index and low black hole mass criteria, and the remaining five do not have sufficient data to determine an accurate \Sersic index or bulge mass. Larger numbers of observed TDEs and more detailed analyses of low concentration post-starburst hosts or high concentration non-bursty host galaxies will be an important test of which mechanisms most affect the TDE rate.


\subsection{Black Hole Binary}
After a galaxy--galaxy merger, the supermassive black hole from each galaxy will inspiral and coalesce. The influence of supermassive black hole binary dynamics on TDEs is the subject of another chapter in this review (Coughlin et al. 2019, ISSI review). We summarize here the relevant points from Coughlin et al. (2019, ISSI review) for the present discussion of the host galaxies.

The TDE rate can be very high (of order 1 per year) for a short ($\sim 1$ Myr) period during coalescence when the secondary black hole approaches the cusp of stars around the primary, at $\sim$pc scale separations. As inspiral continues, the TDE rate will then drop below that of an isolated black hole, and rise once more to a modest rate enhancement of $\sim2-10\times$ that of an isolated black hole once the binary has reached mpc separations. The lightcurves of TDEs can be altered in the case of a tightly bound binary where the debris stream interacts with the companion supermassive black hole. However, if most of the TDEs around black holes in coalescing binaries happens at pc-scale separations, the debris streams will be on significantly smaller scales, and the TDE lightcurves will show no evidence of the companion supermassive black hole. Thus, we are unlikely to see observational effects from the secondary black hole at separations of the same spatial scales which will boost the TDE rate.

The main observable difference between this explanation for the TDE rate enhancement in post-starburst or centrally-concentrated galaxies and the others, is the TDE rate per galaxy. If the TDE rate is very high ($\sim 0.1-1$ per year per galaxy) for a Myr, we should observe a high instance of repeat TDEs per host galaxy, especially over the 10 year run of LSST. The other mechanisms for enhancing the TDE rate described in this section would act over 100 Myr - 1 Gyr, with more modest TDE rates per galaxy, and instances of repeat TDEs would be rare. No repeated TDEs have been observed to date, which means either there are no observed cases where the TDE rate is as high as 1 per several years, or systems hosting multiple TDEs within several years are obscured by dust or otherwise produce a different observational signature than the TDEs discussed in this review.

For now, the likelihood that supermassive black hole binary effects are driving the observed host galaxy distributions can be probed statistically. \citet{French2017} measured the star formation histories for a sample of six post-starburst TDE host galaxies to determine the time since the recent starbursts. If the starburst coincides with the coalescence of the two galaxies, this can also constrain the time since the supermassive black holes started to inspiral on kpc-scales. Most of the TDE host galaxies are less than 600 Myr since starburst. For a secondary to have in-spiraled to pc-scales in 600 Myr, that mass ratio of the two galaxies must be more equal than 12:1 given the dynamical friction timescales. This constrains the TDE rate enhancement in supermassive black hole binaries to be more strongly dependent on the mass ratio than the merger rate, since minor mergers with mass ratios less equal than 12:1 are more common than major mergers. \citet{Stone2018} also argue against the possibility of supermassive black hole binaries causing the observed rate enhancement in post-starburst galaxies by constructing an expected delay time distribution of TDEs after a starburst. Many more TDEs would be expected at times $>1$ Gyr after a starburst, but the observed host galaxies have younger ages. Unless the timescale between the merger and the starburst is fine-tuned, compared to the dynamical friction timescale, the observed host galaxies are not compatible with rate enhancement from a black hole binary scenario.

\subsection{Circumnuclear Gas}
\label{sec:circumnucleargas}

If a supermassive black hole is surrounded by a circumnuclear gas disk, this may act to enhance the TDE rate as stars interact with the disk. \citet{Kennedy2016} predict the TDE rate could be increased by up to $10\times$ by the presence of such a disk. This effect may co-exist with the other mechanisms for affecting the TDE rate discussed here, as mergers are expected to trigger AGN activity \citep[e.g.,][]{Treister2012} as well as bursts of centrally-concentrated star formation. 

However, there are a number of selection effects which may hinder the identification of TDEs in host galaxies with pre-existing circumnuclear gas disks. As discussed by \citet{Law-Smith2017}, \citet{Wevers2019}, and others, observational searches for TDEs will select against AGN host galaxies while trying to avoid classifying AGN flares from variations in the gas accretion rate as TDEs, and many TDEs in AGN host galaxies will be heavily extincted by dust. One possibility for identifying TDEs in AGN host galaxies with heavy dust obscuration is through multi-epoch radio observations. \citet{Mattila2018} identify a TDE in Arp 299, a Seyfert II galaxy and dust-obscured LIRG \citep{Pereira-Santaella2015}, via a growing radio jet in one of the two nuclei in this merging system. Identifying the accretion of an individual star in an AGN with a high gas accretion rate may prove unfeasible, or only possible to assess statistically.

\subsection{Other Dynamical or Secular Effects}

There are several other effects that might increase the TDE rate in post-merger galaxies, causing the observed host galaxy preferences. A more complete description of these effects can be found in the chapter of this review by Stone et al. (2020, ISSI review); we very briefly mention two mechanisms here. First, a radial anisotropy of the orbits after a merger would increase the rate of TDEs. \citet{Stone2018} modeled the dependence of the rate enhancement on the time since the starburst, parameterizing the radial anisotropy as $\beta \equiv 1- \frac{T_\perp}{2 T_{\|}}$, where $T_\perp$ and $T_{\|}$ are the kinetic energies of the tangential and radial motion, respectively\footnote{$\beta$ is defined such that if all orbits are purely radial, $\beta=1$, and if all orbits are purely tangential, $\beta = -\infty$.}. \citet{Stone2018} find that the observed host galaxies could be explained if  $\beta > 0.6$. A triaxial nuclear potential \citep{Merritt2004} could have a similar effect of enhancing the TDE rate after a nuclear starburst. Second, \citet{Madigan2018} predict an enhanced TDE rate that declines with time after a nuclear starburst due to the formation of an eccentric stellar disk, where the TDE rate would be increased due to secular effects and could potentially reach 0.1 -- 1 gal$^{-1}$ yr$^{-1}$. However, the timescale over which this effect might produce a high TDE rate is uncertain. The high spatial and spectral resolution of next-generation 30-m class telescopes will provide crucial tests of these mechanisms in nearby TDE hosts.

\section{Implications for TDE Rates}
\label{sec:rates}

\begin{figure*}
\includegraphics[width=0.5\textwidth]{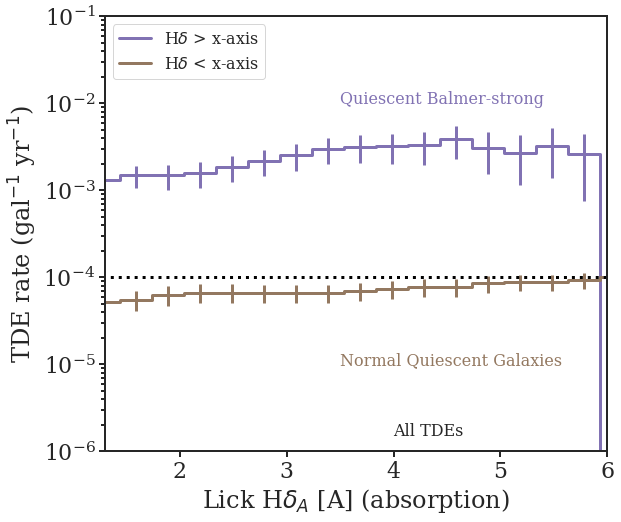}
\includegraphics[width=0.5\textwidth]{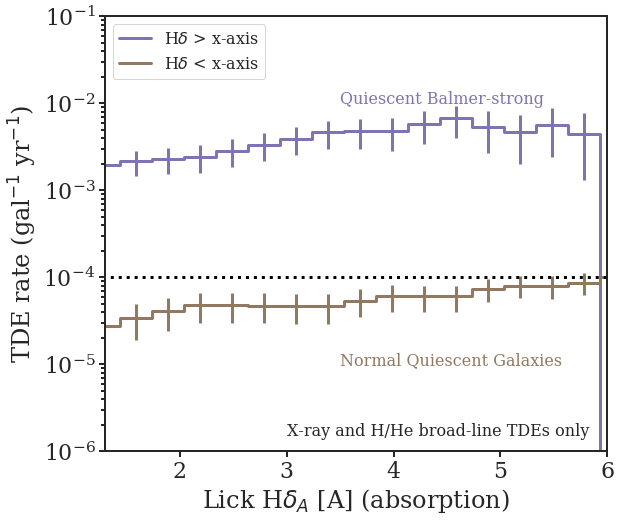}
\caption{TDE rate (scaled to a fiducial value of $10^{-4}$ per galaxy per year as measured by e.g., \citealt{vanvelzen2018}) in quiescent galaxies above and below each value of H$\delta$ absorption. While the rate enhancement in quiescent Balmer-strong and post-starburst galaxies results in higher TDE rates, the resulting suppression of TDE rates in quiescent galaxies that do no meet a given H$\delta$ absorption threshold is mild, at $<3\times$.}
\label{fig:rate_change}
\end{figure*}

The variation in TDE rates by host galaxy type has implications for the relative TDE rates in different types of galaxies. The TDE rate averaged across all galaxy types is observed to be $\sim 10^{-4}$ per galaxy per year \citep{vanvelzen2018}. We quantify the TDE rates in quiescent Balmer-strong and post-starburst galaxies compared to normal quiescent galaxies in Figure \ref{fig:rate_change}, and how this depends on the definition of ``Balmer-strong". The TDE rate in quiescent Balmer-strong and post-starburst galaxies is 1-3$\times10^{-3}$ per galaxy per year, depending on the threshold used to define such galaxies. An enhancement in the TDE rate in certain galaxies will necessarily result in a decreased TDE rate in other galaxies given a measurement of the total TDE rate. Is this lowered rate enough to cause tension with theoretical predictions for the TDE rate in normal quiescent galaxies ($\sim$ few$\times 10^{-4}$ per galaxy per year, e.g., \citealt[]{Stone2016b})? Depending on the definition of ``Balmer-strong" vs. ``normal" quiescent galaxies, the TDE rate in normal quiescent galaxies is $0.3-1\times10^{-4}$ per galaxy per year. We stress here that despite the enhanced TDE rates observed in post-starburst and quiescent Balmer-strong galaxies, the number of TDEs in normal galaxies is still high enough to avoid a crisis in their TDE rates. The large number of galaxies which do not meet the quiescent Balmer-strong or post-starburst selection compensates for the lower TDE rate in such ``normal" galaxies. This rate suppression of $\lesssim3\times$ in normal quiescent galaxies is comparable to the uncertainties in predicting the theoretical TDE rate and determining the observed TDE rate in various surveys.

\section{Using the Host Galaxy Information in Transient Surveys}
\label{sec:surveys}

\subsection{Identifying TDEs using a priori Host Galaxy Information}

In addition to being used to understand what drives the TDE rates, the unique properties of TDE host galaxies may be useful for selecting candidate TDEs for spectroscopic follow-up observations in large transient surveys. The large volume of transient alerts faced by current and future surveys such as ZTF and LSST means that not all transients can be followed up for spectroscopic confirmation and further study with triggered space-based observations and high cadence photometry. Methods to flag likely TDEs will be essential for future TDE studies. One method to flag likely TDE candidates is to use a priori information about their likely host galaxies. Transients discovered in pre-identified likely TDE host galaxies, based on (for example) E+A spectroscopic signatures or concentration indices, can then be systematically classified with dedicated spectroscopic observations to efficiently select TDEs for further photometric and spectroscopic monitoring.

For transient surveys in the Northern hemisphere where transient detections will have a significant overlap with large spectroscopic surveys (especially SDSS), the detailed properties of likely host galaxies can be used to predict which candidate detections are likely TDEs, supernovae, or other transient phenomenon, and flag interesting objects for follow-up. However, a significant portion of the LSST footprint will not be covered by large spectroscopic surveys, so photometric criteria will be useful. 

A technique of selecting likely TDEs for followup based on the colour (and thus star formation rate) of the host galaxy is used by iPTF/ZTF \citep{Hung2018} in order to reduce contamination by supernovae. Further cuts on the recent star formation history or concentration of host galaxies may further refine this selection, independent of the physical reason for the enhanced TDE rate in such galaxies.

\citet{French2018} present a method for identifying likely TDE candidates using photometrically-identified quiescent Balmer-strong galaxies. A Random Forest classifier trained on spectroscopically-identified quiescent Balmer-strong galaxies can be used to detect such galaxies in LSST using LSST photometry in addition to archival photometry from {\it GALEX} and {\it WISE}. Because these galaxies have low star formation rates and thus low supernova rates, contamination from other transients will be low. 

Other possible methods for identifying likely TDE hosts can use photometric information on the light concentration, given by either the \Sersic index or bulge fraction that may predict a TDE overabundance in a related set of likely host galaxies (see discussion in \S\ref{sec:rate_conc}). For instance, choosing nuclear transients in high-\Sersic galaxies could significantly increase the success of confirming TDEs.

There are several important benefits to selecting TDEs for spectroscopic follow-up. In addition to identifying a large number of events, this method can identify TDEs with properties that diverge from the known set of TDEs. The small number of optical/UV bright TDEs studied to date already show a large variety in their peak magnitude, colour evolution, and decline timescale \citep{Holoien2016, Blagorodnova2017, Wevers2017}. Furthermore, a priori selection of the likely host galaxies means candidate events can be followed-up early, even before a light curve is obtained. Early time light curve information will enable more detailed modeling, and may be able to be used to measure black hole masses \citep{Mockler2018}.

Any method for selecting transient events for follow-up based on the host galaxy properties will bias the host galaxy properties of the resulting sample. Thus, these selection methods are complementary to other methods which are un-targeted or use information on the transient properties alone.

\section{Galaxy Studies Using TDEs}
\label{sec:galaxystudies}
Current and upcoming transient surveys, such as LSST in the optical and e-ROSITA in the X-ray, are expected to detect hundreds to thousands of TDEs each year \citep{VanVelzen2011, Khabibullin2014}. In light of the theoretical predictions and observational evidence that these TDEs should occur in the lowest mass galaxies still harbouring black holes \citep{Wang2004}, TDEs could become beacons for studying the black hole mass functions and spin distributions of quiescent supermassive black holes. 

While measuring velocity dispersions of large samples of increasingly faint galaxies will become prohibitively expensive, alternatives are emerging to measure the black hole mass. For example, it is expected that the shape of a TDE lightcurve depends mainly on the mass of the black hole and the mass of the disrupted star. \citet{Mockler2018} exploit this by using lightcurve models to infer the black hole mass under some assumptions for the mass of the disrupted star. They show that this method yields similar uncertainties when compared to other ways of estimating the black hole mass (e.g. galaxy scaling relations), while further improvements (such as including realistic models for the stellar structure) could decrease potential systematic effects. This can provide the opportunity to measure black hole masses of galaxies that are too faint for velocity dispersion measurements, as well as large samples of galaxies, as long as the lightcurve is well sampled around peak and ideally in more then one filter. 

These mass measurements will be invaluable for TDE demographics studies, but can also potentially contribute to other areas of galaxy evolution, such as the cosmic growth of SMBHs in quiescent galaxies and the BH occupation fraction and mass function of dwarf galaxies. On the high end of the mass function, observed TDE rates will be affected by the presence of an event horizon cutoff, as a consequence of the tidal radius becoming smaller than the event horizon and making disruptions invisible to outside observers. A main sequence star disrupted by a Schwarzschild black hole of mass $\ge 10^8 M_\odot$ will not produce an observable flare. However, the spin of the SMBH will affect this cutoff mass, such that TDEs would still be observable around $10^8-10^9 M_\odot$ SMBHs at high ($a\sim0.9-0.999$) spins \citep{Kesden2012, Leloudas2016} A large number of TDEs with black hole masses $\ge 10^8 M_\odot$ would imply a large fraction of supermassive black holes with high spin parameters. The spin distribution for supermassive black holes has implications for their accretion and merger history, as coherent gas accretion is expected to spin up black holes while frequent mergers will spin them down \citep[e.g.,][]{Volonteri2003, Hughes2003}. The effect of the spin distribution on the observed TDE rate is discussed further by Stone et al. (2020, ISSI review). The possibilities for observing TDEs in dwarf galaxies with intermediate mass black holes are discussed by Maguire et al. (2020, ISSI review).

\section{Summary and Discussion}
\label{sec:summary}
We have summarized the observed host galaxy properties of X-ray bright and optical/UV bright TDEs and discussed the possible physical mechanisms that could drive the observed correlations between host galaxy properties and the TDE rate. While the black hole masses of TDEs have similar distributions between X-ray and optical/UV bright TDEs, compiling a large sample of known TDEs allows us to identify a significant shift in the stellar masses and absolute magnitude distributions between the two samples, with optical/UV TDEs having higher stellar masses than the X-ray TDEs. This may be due to an underlying trend in black hole mass we do not yet have the statistical power to resolve. Future work on determining the black hole masses of new TDEs and understanding the selection effects of TDEs identified in the X-ray or optical will be needed.

Most TDE host galaxies are quiescent, with little current star formation, and TDEs are over-represented in galaxies with post-starburst or quiescent Balmer-strong star formation histories. We present new estimates of the TDE rate enhancement in these samples depending on the definition of these classes and the types of TDEs considered. For the TDE host galaxies considered in this review, 5/41 (12\%) are post-starburst galaxies and 13/41 (32\%) are either quiescent Balmer-strong or post-starburst. Of the 4 X-ray TDEs, 3 (75\%) are quiescent Balmer-strong and 1 (25\%) is post-starburst. Of the 15 broad H/He line TDEs, 9 (60\%) are quiescent Balmer-strong and 5 (33\%) are post-starburst. We note again that these TDE classifications are tentative and subject to a number of observational biases. More observations of the host galaxies for a large sample of well-characterized TDEs are needed to overcome these uncertainties and the small-number statistics limiting our precision here. While controlling for some galaxy properties, notably bulge color and central concentration, can reduce the TDE enhancement rate in post-starburst or quiescent Balmer-strong galaxies, these properties are strongly correlated with galaxy SFH during the post-starburst phase. The TDE enhancement rates in such hosts are thus consistent (given the small number statistics) between the studies of \citet{French2016}, \citet{Law-Smith2017} and \citet{Graur2018}. While the rate enhancement in quiescent Balmer-strong and post-starburst galaxies results in higher TDE rates, the resulting suppression of TDE rates in normal quiescent galaxies is mild, at $<3\times$.

TDE host galaxies have higher concentrations of stellar light than expected given their stellar mass or black hole mass. We aggregate here observations from many sources of TDE host galaxy optical light concentrations, and present new versions of two analyses from the literature. We consider the effect of volume-correcting the SDSS comparison sample on the result by \citet{Law-Smith2017} that TDE host galaxies have high \Sersic indices for their black hole masses, finding that despite the large number of low black hole mass galaxies inferred from the volume correction, the TDE hosts still lie at high \Sersic indices for their black hole masses, compared with the rest of the galaxy sample. We also add measurements for two additional TDE host galaxy velocity dispersions from \citet{Wevers2019} to the analysis of \citet{Graur2018} that shows TDE hosts have high stellar surface densities for their velocity dispersions, and find the conclusions unchanged.

The ionisation states of TDE host galaxies span a range from star-formation dominated to AGN-dominated, and we identify many TDE host galaxies with signs of on-going gas accretion by the SMBH. However, given the significant selection biases affecting this distribution, and the uncertainties in identifying especially low-luminosity AGN or LINERs, we urge caution in its interpretation.  The extra-galactic environments of TDE host galaxies show no signs of being different than the general galaxy population or the post-starburst galaxy population, but this comparison is especially limited by the small numbers of TDE hosts with well-studied extra-galactic environments.

We summarize the state of several possible explanations for the links between the TDE rate and host galaxy type, including the effect of stellar overdensities, black hole binaries, circumnuclear gas, and dynamical or secular effects. We present estimates of the TDE rate for different host galaxy types and quantify the degree to which rate enhancement in some groups results in rate suppression in others. We discuss the possibilities for using TDE host galaxies to assist in identifying TDEs in upcoming large transient surveys and possibilities for TDE observations to be used to study their host galaxies.

We note that the TDEs considered here are a relatively small number of events, with classifications subject to observational and selection biases which may change in future studies of TDEs.

We identify the following as important open questions in this field: 
\begin{enumerate}
    \item What is the primary driver of the observed correlations between galaxy-scale host properties and the TDE rate?
    \item What are the distributions of stellar mass and the stellar kinematics of TDE host galaxies within the radius of gravitational influence of their supermassive black holes?
    \item Do the same physical effects drive the observed enhanced TDE rates in post-starburst or quiescent Balmer-strong galaxies and galaxies with high central concentrations?
    \item Are the high central concentrations at kpc scales of optical stellar light seen in TDE hosts correlated with high central concentrations of stars at the nucleus?
    \item What is the unobscured TDE rate in starburst galaxies? Does the enhancement during the post-starburst phase arise from a higher absolute rate during this phase or simply a higher observed rate due to dust obscuration during the starburst phase and/or selection bias against AGN in starburst galaxies?
    \item How does the presence of an existing AGN of varying accretion rates affect the rates and observability of TDEs?
    \item What is the connection between TDEs and Narrow-Line Seyfert Is?
    \item What is the connection between TDEs and the LINER-like emission present in most post-starburst or quiescent Balmer-strong galaxies? What can cases where a host galaxy has pre-TDE spectroscopy tell us about the evolution of narrow line emission from TDEs vs. AGN?
    \item Are there differences in the host galaxies depending on the type of TDE?
    \item What drives the observed difference in stellar mass between the optical and X-ray bright TDEs? Is the difference due to a difference in intrinsic black hole mass? 
    Are the observed trends affected by selection bias in how TDEs are identified in the optical vs. X-ray? 
    \item How can we best use our understanding of TDE host galaxies to study supermassive black holes and their host galaxies using new observations from next-generation surveys like LSST and \textit{eROSITA}? How can we best compare the samples of TDEs discovered using LSST vs. \textit{eROSITA}? 
\end{enumerate}

Planned time-domain programs in the next decade will discover hundreds to thousands of new TDEs and will enable the detailed study of TDEs and their host galaxies with significantly greater statistical power.

\begin{acknowledgements}

We thank the referees for their detailed feedback and helpful suggestions, which have improved this review.

The authors thank ISSI for their support and hospitality and the review organizers for their leadership in coordinating this set of reviews.

K.D.F. is supported by Hubble Fellowship grant HST-HF2-51391.001-A, provided by NASA through a grant from the Space Telescope Science Institute (STScI), which is operated by the Association of Universities for Research in Astronomy, Inc., under NASA contract NAS5-26555. A.I.Z. acknowledges support from NASA through STScI grant HST-GO-14717.001-A. T.W. is funded in part by European Research Council grant 320360 and by European Commission grant 730980. O.G. is supported by an NSF Astronomy and Astrophysics Fellowship under award AST-1602595.

Funding for SDSS-III has been provided by the Alfred P. Sloan Foundation, the Participating Institutions, the National Science Foundation, and the U.S. Department of Energy Office of Science. The SDSS-III website is http://www.sdss3.org/. SDSS-III is managed by the Astrophysical Research Consortium for the Participating Institutions of the SDSS-III Collaboration, including the University of Arizona, the Brazilian Participation Group, Brookhaven National Laboratory, University of Cambridge, Carnegie Mellon University, University of Florida, the French Participation Group, the German Participation Group, Harvard University, the Instituto de Astrofisica de Canarias, the Michigan State/Notre Dame/JINA Participation Group, Johns Hopkins University, Lawrence Berkeley National Laboratory, Max Planck Institute for Astrophysics, Max Planck Institute for Extraterrestrial Physics, New Mexico State University, New York University, the Ohio State University, Pennsylvania State University, University of Portsmouth, Princeton University, the Spanish Participation Group, University of Tokyo, University of Utah, Vanderbilt University, University of Virginia, University of Washington, and Yale University.

This publication makes use of data products from the Wide-field Infrared Survey Explorer \citep{2010AJ....140.1868W}, which is a joint project of the University of California, Los Angeles, and the Jet Propulsion Laboratory/California Institute of Technology, funded by the National Aeronautics and Space Administration.

This work made use of the IPython package \citep{PER-GRA:2007}. This research made use of SciPy \citep{jones_scipy_2001}. This research made use of Astropy, a community-developed core Python package for Astronomy \citep{2013A&A...558A..33A}. This research made use of NumPy \citep{van2011numpy}. 
\end{acknowledgements}

\bibliographystyle{aps-nameyear}      
\bibliography{extra_hostgal.bib}

\end{document}

%% file: tde_info.txt
ASASSN14li$^c$  & 12:48:15.23 &	+17:46:26.44 & 0.02058 &  X-ray TDE & 1  & 0\\
  Swift J1644                    & 16:44:49.30	& +57:34:51.00 & 0.3534    & X-ray TDE & 0  & 0\\
  XMM J0740                      & 07:40:08.09 &	$-$85:39:31.30 & 0.0173   & X-ray TDE & 0  & 0\\ 
  ASASSN15oi  & 20:39:09.18	& $-$30:45:20.10 & 0.0484   & X-ray TDE & 1  & 0\\
  SDSS J1201$^c$                   &   12:01:36.03 &	+30:03:05.52 & 0.146   & Likely X-ray TDE & 0  & 0\\
  2MASX J0249 & 02:49:17.32	& $-$04:12:52.20 & 0.0186    & Likely X-ray TDE  & 0  & 0\\
  PTF10iya                     & 14:38:40.98 &	+37:39:33.45 & 0.22405  & Likely X-ray TDE  & 0  & 0\\
  SDSS J1311                   & 13:11:22.15	& $-$01:23:45.61 & 0.195    & Likely X-ray TDE  & 0  & 0\\
  SDSS J1323$^c$                   & 13:23:41.97 &	+48:27:01.26 & 0.0875   & Likely X-ray TDE  & 0  & 0\\
  3XMM J1521                   &   15:21:30.73 &	+07:49:16.52 & 0.17901   & Likely X-ray TDE & 0  & 0\\
  3XMM J1500                   & 15:00:52.07 & +01:54:53.82 & 0.145    & Likely X-ray TDE  & 0  & 0\\
 PS18kh & 07:56:54.53 & +34:15:43.63 & 0.074 & Likely X-ray TDE & 1  & 0\\
  RX J1242-A  &  12:42:38.54 & $-$11:19:20.85   & 0.05  & Possible X-ray TDE  & 0  & 0\\
  RX J1242-B  & 12:42:38.16 & $-$11:19:13.62 & 0.05      & Possible X-ray TDE  & 0  & 0\\
  RX J1420-A                   & 14:20:24.39 & +53:34:11.14 & 0.148     & Possible X-ray TDE  & 0  & 0\\
  RX J1420-B                   & 14:20:24.52 & +53:34:15.72 & 0.147     & Possible X-ray TDE & 0  & 0\\   
  SDSS J0159                   & 01:59:57.64 &	+00:33:10.49 & 0.31167  & Possible X-ray TDE  & 0  & 0\\
  RBS 1032$^c$                     & 11:47:26.80 &	+49:42:59.00 & 0.02604   & Possible X-ray TDE  & 0  & 0\\
  RX J1624                     &   16:24:56.66 &	+75:54:56.09 & 0.0636    & Possible X-ray TDE  & 0  & 0\\
  NGC 5905                     & 15:15:23.32 &	+55:31:01.59  & 0.01131  & Possible X-ray TDE  & 0  & 0\\
GALEX D3-13 & 14:19:29.81 & +52:52:06.37 & 0.3698 & Possible X-ray TDE  & 0  & 0\\
  iPTF16fnl       &   00:29:57.01 & +32:53:37.24 & 0.0163   & Optical/UV TDE & 1  & 0\\
  PS16dtm$^c$                           & 01:58:04.75 & $-$00:52:21.87 & 0.0804    & Optical/UV TDE & 0  & 0\\
  F01004                           & 01:02:50.01 & $-$22:21:57.22 & 0.1178   & Optical/UV TDE  & 0  & 0\\
  GALEX D23H-1                           & 23:31:59.54 & +00:17:14.58 & 0.1855    & Optical/UV TDE  & 0  & 0\\
  PS1-11af                         &   09:57:26.82 & +03:14:00.94 & 0.4046    & Optical/UV TDE  & 0  & 0\\
  SDSS J0952$^c$      & 09:52:09.56 &	+21:43:13.24 & 0.0789   & Optical/UV TDE  & 0  & 1\\
  SDSS J1342$^c$     & 13:42:44.42	& +05:30:56.14 & 0.0366   & Optical/UV TDE & 0  & 1\\
  SDSS J1350$^c$      & 13:50:01.51 & +29:16:09.71 & 0.0777   & Optical/UV TDE  & 0  & 1\\
  SDSS TDE1                             &   23:42:01.41 & +01:06:29.30 & 0.1359   & Optical/UV TDE   & 0  & 0\\
  SDSS TDE2            & 23:23:48.62	& $-$01:08:10.34 & 0.2515    & Optical/UV TDE & 1  & 0\\
  SDSS J0748$^c$  & 07:48:20.67 & +47:12:14.23 & 0.0615     & Optical/UV TDE & 1  & 1\\
  ASASSN14ae$^c$    & 11:08:40.12	& +34:05:52.23 & 0.0436   & Optical/UV TDE & 1 & 0 \\
  PTF09axc      & 14:53:13.08	& +22:14:32.27 & 0.1146     & Optical/UV TDE & 1  & 0\\
  PTF09djl      & 16:33:55.97	& +30:14:16.65 & 0.184      & Optical/UV TDE & 1  & 0\\
  PTF09ge$^c$       & 14:57:03.18	& +49:36:40.97 & 0.064     & Optical/UV TDE & 1  & 0\\
  PS1-10jh      & 16:09:28.28	& +53:40:23.99 & 0.1696     & Optical/UV TDE & 1  & 0\\
  iPTF15af$^c$       & 08:48:28.12 &  +22:03:33.58 & 0.079      & Optical/UV TDE & 1 & 0 \\
  iPTF16axa                        & 17:03:34.36 &  +30:35:36.8 & 0.108     & Optical/UV TDE & 0  & 0\\
 GALEX D1-9 &  02:25:17.00 & $-$04:32:59.00 & 0.326 & Optical/UV TDE & 0  & 0\\
AT2018dyk$^c$ & 15:33:08.02 & +44:32:08.20 & 0.037  & Optical/UV TDE & 1  & 0\\
 AT2018bsi & 08:15:26.62 & +45:35:31.95 & 0.051 & Optical/UV TDE & 1   & 0\\
 ASASSN18zj$^c$ & 10:06:50.74 & +01:41:34.37 & 0.046 & Optical/UV TDE & 1 & 0 \\

%% file: tde_table_21.txt
\\[-\normalbaselineskip]
\multicolumn{5}{c}{X-ray TDEs} \\
\hline
ASASSN14li$\dagger$	&	10.6	&	-18.8	&	81(2)	&	6.23$^{+0.39}_{-0.40}$	\\
Swift J1644	&	-	&	-	&	-	&	-	\\
XMM J0740	&	-	&	-	&	-	&	-	\\
ASASSN15oi$\dagger$	&	9.9	&	-19.3	&	61(7)	&	5.71$^{+0.60}_{-0.57}$	\\
\hline
\multicolumn{5}{c}{Likely X-ray TDEs} \\
\hline
SDSS J1201	&	10.4	&	-20.6	&	122(4)	&	7.18$^{+0.41}_{-0.41}$	\\
2MASX J0249	&	9.1	&	-17.5	&	43(4)	&	4.93$^{+0.55}_{-0.53}$	\\
PTF10iya	&	9.3	&	-20.0	&	-	&	-	\\
SDSS J1311	&	8.7	&	-18.6	&	-	&	-	\\
SDSS J1323	&	9.8	&	-18.9	&	75(4)	&	6.15$^{+0.46}_{-0.45}$	\\
3XMM J1521	&	9.9	&	-19.2	&	58(2)	&	5.61$^{+0.41}_{-0.41}$	\\
3XMM J1500	&	9.3	&	-19.1	&	59(3)	&	5.64$^{+0.45}_{-0.45}$	\\
PS18kh$\dagger$	&	9.6	&	-19.0	&	-	&	-	\\
\hline
\multicolumn{5}{c}{Possible X-ray TDEs} \\
\hline
RX J1242-A	&	10.3	&	-21.0	&	-	&	-	\\
RX J1242-B	&	-	&	-	&	-	&	-	\\
RX J1420-A	&	10.3	&	-20.3	&	131(13)	&	7.33$^{+0.56}_{-0.54}$	\\
RX J1420-B	&	-	&	-	&	-	&	-	\\
SDSS J0159	&	10.7	&	-21.8	&	124(10)	&	7.21$^{+0.52}_{-0.50}$	\\
RBS 1032	&	9.0	&	-17.7	&	49(7)	&	5.25$^{+0.67}_{-0.62}$	\\
RX J1624	&	10.4	&	-20.8	&	155(9)	&	7.68$^{+0.45}_{-0.45}$	\\
NGC 5905	&	10.0	&	-20.2	&	97(5)	&	6.69$^{+0.45}_{-0.44}$	\\
GALEX D3-13	&	10.7	&	-20.8	&	133(6)	&	7.36$^{+0.43}_{-0.44}$	\\
\hline
\multicolumn{5}{c}{Optical/UV TDEs} \\
\hline
iPTF16fnl$\dagger$	&	9.8	&	-19.8	&	55(2)	&	5.50$^{+0.42}_{-0.42}$	\\
PS16dtm	&	9.6	&	-19.3	&	45(13)	&	5.07$^{+0.88}_{-1.06}$	\\
F01004	&	9.8	&	-21.0	&	132(29)	&	7.34$^{+0.76}_{-0.86}$	\\
GALEX D23H-1	&	10.3	&	-20.1	&	84(4)	&	6.39$^{+0.44}_{-0.44}$	\\
PS1-11af	&	10.1	&	-20.1	&	-	&	-	\\
SDSS J0952$^*$	&	10.0	&	-20.2	&	-	&	-	\\
SDSS J1342$^*$	&	9.5	&	-19.0	&	72(6)	&	6.06$^{+0.51}_{-0.52}$	\\
SDSS J1350$^*$	&	-	&	-	&	-	&	-	\\
SDSS TDE1	&	10.1	&	-19.2	&	126(7)	&	7.25$^{+0.45}_{-0.46}$	\\
SDSS TDE2$\dagger$	&	10.6	&	-20.6	&	-	&	-	\\
SDSS J0748$\dagger$$^*$	&	9.9	&	-20.0	&	126(7)	&	7.25$^{+0.45}_{-0.46}$	\\
ASASSN14ae$\dagger$	&	10.8	&	-19.2	&	53(2)	&	5.42$^{+0.46}_{-0.46}$	\\
PTF09axc$\dagger$	&	10.0	&	-20.2	&	60(4)	&	5.68$^{+0.48}_{-0.49}$	\\
PTF09djl$\dagger$	&	10.1	&	-20.0	&	64(7)	&	5.82$^{+0.56}_{-0.58}$	\\
PTF09ge$\dagger$	&	10.1	&	-19.5	&	82(2)	&	6.31$^{+0.39}_{-0.39}$	\\
PS1-10jh$\dagger$	&	9.5	&	-18.1	&	65(3)	&	5.85$^{+0.44}_{-0.44}$	\\
iPTF15af$\dagger$	&	10.2	&	-18.0	&	106(2)	&	6.88$^{+0.38}_{-0.38}$	\\
iPTF16axa	&	10.1	&	-19.4	&	82(3)	&	6.34$^{+0.42}_{-0.42}$	\\
AT2018dyk$\dagger$	&	10.6	&	-21.4	&	112(4)	&	7.00$^{+0.41}_{-0.42}$	\\
AT2018bsi$\dagger$	&	10.3	&	-20.8	&	-	&	-	\\
ASASSN18zj$\dagger$	&	9.5	&	-19.1	&	60(5)	&	5.68$^{+0.51}_{-0.52}$	\\

%% file: tde_table_22.txt
\\[-\normalbaselineskip]
\multicolumn{7}{c}{X-ray TDEs} \\
\hline
ASASSN14li$\dagger$  &  -0.6  &  0.5  &  5.7  &  0.6  &  0.01  &  PSB  \\ 
Swift J1644  &  -2.5  &  0.8  &  4.7  &  1.1  &  --  &  QBS  \\ 
XMM J0740  &  -0.3  &  0.6  &  0.4  &  0.4  &  --  &  Quiescent  \\ 
ASASSN15oi$\dagger$  &  0.1  &  0.3  &  1.9  &  0.7  &  --  &  QBS  \\ 
\hline
\multicolumn{7}{c}{Likely X-ray TDEs} \\
\hline
SDSS J1201  &  0.7  &  0.3  &  -1.1  &  2.4  &  --  &  Quiescent  \\ 
2MASX J0249  &  -5.7  &  0.6  &  0.4  &  0.5  &  --  &  SF  \\ 
PTF10iya  &  -20.5  &  0.6  &  2.9  &  0.9  &  --  &  SF  \\ 
SDSS J1311  &  -2.1  &  1.5  &  3.6  &  1.1  &  --  &  QBS  \\ 
SDSS J1323  &  -0.2  &  0.5  &  -1.2  &  1.5  &  0.01  &  Quiescent  \\ 
3XMM J1521  &  0.8  &  1.1  &  -1.5  &  2.4  &  --  &  Quiescent  \\ 
3XMM J1500  &  -45.1  &  0.8  &  1.4  &  1.7  &  --  &  SF  \\ 
\hline
\multicolumn{7}{c}{Possible X-ray TDEs} \\
\hline
RX J1242-A  &  1.1  &  0.8  &  0.9  &  1.2  &  --  &  Quiescent  \\ 
RX J1242-B  &  -0.9  &  0.9  &  -0.4  &  2.7  &  --  &  Quiescent  \\ 
RX J1420-A  &  -0.2  &  0.9  &  -2.8  &  2.0  &  0.03  &  Quiescent  \\ 
RX J1420-B  &  -71.6  &  1.7  &  5.8  &  3.4  &  0.54  &  SF  \\ 
SDSS J0159  &  -19.9  &  0.8  &  1.7  &  1.0  &  5.25  &  SF  \\ 
RBS 1032  &  -0.5  &  0.4  &  4.1  &  0.4  &  0.0  &  QBS  \\ 
RX J1624  &  0.6  &  1.3  &  -1.1  &  2.1  &  --  &  Quiescent  \\ 
NGC 5905  &  -28.4  &  0.1  &  --  &  --  &  --  &  SF  \\ 
\hline
\multicolumn{7}{c}{Optical/UV TDEs} \\
\hline
iPTF16fnl$\dagger$  &  0.8  &  0.6  &  5.8  &  0.3  &  --  &  PSB  \\ 
PS16dtm  &  -31.8  &  0.4  &  -0.2  &  1.1  &  0.26  &  SF  \\ 
F01004  &  -47.0  &  0.2  &  -0.2  &  0.8  &  --  &  SF  \\ 
GALEX D23H-1  &  -13.3  &  0.9  &  4.3  &  1.5  &  --  &  SF  \\ 
PS1-11af  &  0.7  &  1.2  &  1.5  &  1.4  &  --  &  QBS  \\ 
SDSS J0952$^*$  &  -27.8  &  0.4  &  -2.0  &  1.1  &  1.6  &  SF  \\ 
SDSS J1342$^*$  &  -15.1  &  0.5  &  -1.0  &  1.3  &  0.07  &  SF  \\ 
SDSS J1350$^*$  &  -20.3  &  0.4  &  0.9  &  1.3  &  0.55  &  SF  \\ 
SDSS TDE1  &  1.2  &  1.0  &  -1.3  &  1.3  &  --  &  Quiescent  \\ 
SDSS TDE2$\dagger$  &  -4.5  &  0.5  &  3.7  &  0.6  &  --  &  SF  \\ 
SDSS J0748$\dagger$$^*$  &  -11.4  &  1.0  &  1.2  &  0.8  &  0.87  &  SF  \\ 
ASASSN14ae$\dagger$  &  -0.7  &  0.4  &  3.4  &  0.8  &  0.01  &  QBS  \\ 
PTF09axc$\dagger$  &  -1.1  &  0.7  &  4.9  &  0.4  &  --  &  PSB  \\ 
PTF09djl$\dagger$  &  -0.3  &  0.7  &  4.7  &  0.5  &  --  &  PSB  \\ 
PTF09ge$\dagger$  &  -1.7  &  0.8  &  0.3  &  0.7  &  0.05  &  Quiescent  \\ 
PS1-10jh$\dagger$  &  -0.5  &  0.7  &  1.7  &  0.8  &  --  &  QBS  \\ 
iPTF15af$\dagger$  &  -1.7  &  0.3  &  1.3  &  1.9  &  0.02  &  QBS  \\ 
iPTF16axa  &  -1.1  &  1.7  &  0.2  &  1.5  &  --  &  Quiescent  \\ 
PS18kh  &  -0.11  &  0.47  &  0.16  &  0.56  &  --  &  Quiescent  \\ 
AT2018dyk$\dagger$  &  -2.17  &  0.11  &  -0.27  &  0.48  &  2.54  &  Quiescent  \\ 
AT2018bsi$\dagger$  &  -5.25  &  0.69  &  -0.73  &  0.47  &  --  &  SF  \\ 
ASASSN18zj$\dagger$  &  -0.35  &  0.14  &  5.43  &  0.41  &  0.01  &  PSB  \\ 

%% file: tde_table_23.txt
\\[-\normalbaselineskip]
\multicolumn{6}{c}{X-ray TDEs} \\
\hline
ASASSN-14li$\dagger$  &  4.91  &  6.17  &  0.015  &  ${10.1}_{-0.2}^{+0.2}$  &  63 $\pm$ 3  \\ 
\hline
\multicolumn{6}{c}{Likely X-ray TDEs} \\
\hline
SDSSJ1201  &  5.61  &  --  &  0.008  &  ${9.5}_{-0.3}^{+0.2}$  &  122 $\pm$ 4  \\ 
SDSSJ1323  &  5.03  &  6.52  &  0.002  &  ${9.5}_{-0.2}^{+0.2}$  &  75 $\pm$ 10  \\ 
3XMM J1521  &  --  &  --  &  --  &  ${9.5}_{-0.3}^{+0.2}$  &  62 $\pm$ 14  \\ 
\hline
\multicolumn{6}{c}{Possible X-ray TDEs} \\
\hline
RX J1420-A  &  --  &  --  &  --  &  ${9.7}_{-0.2}^{+0.2}$  &  131 $\pm$ 13  \\ 
RX J1420-B  &  --  &  --  &  --  &  ${7.4}_{-0.3}^{+0.3}$  &  168 $\pm$ 52  \\ 
SDSSJ0159  &  --  &  --  &  --  &  $< 9.9$  &  128 $\pm$ 17  \\ 
RBS1032  &  2.24  &  5.62  &  0.018  &  ${9.5}_{-0.2}^{+0.2}$  &  36 $\pm$ 9  \\ 
NGC 5905  &  --  &  --  &  --  &  ${10.1}_{-0.2}^{+0.3}$  &  97 $\pm$ 5  \\ 
GALEX D3-13  &  --  &  --  &  --  &  ${9.4}_{-0.2}^{+0.2}$  &  133 $\pm$ 6  \\ 
\hline
\multicolumn{6}{c}{Optical/UV TDEs} \\
\hline
iPTF16fnl$\dagger$  &  --  &  --  &  --  &  ${9.6}_{-0.3}^{+0.3}$  &  55 $\pm$ 2  \\ 
PS16dtm  &  --  &  --  &  --  &  ${9.4}_{-0.2}^{+0.2}$  &  45 $\pm$ 13  \\ 
GALEX D23-H1  &  --  &  --  &  --  &  ${9.2}_{-0.2}^{+0.2}$  &  86 $\pm$ 14  \\ 
SDSSJ0952$^*$  &  7.98  &  7.04  &  0.011  &  $< 10.4$  &  --  \\ 
SDSSJ1342$^*$  &  2.56  &  6.51  &  0.012  &  ${9.7}_{-0.2}^{+0.3}$  &  72 $\pm$ 6  \\ 
SDSSJ1350$^*$  &  4.57  &  7.47  &  0.003  &  ${9.3}_{-0.3}^{+0.3}$  &  --  \\ 
SDSS TDE1  &  --  &  --  &  --  &  ${9.5}_{-0.2}^{+0.2}$  &  137 $\pm$ 12  \\ 
SDSSJ0748$\dagger$$^*$  &  1.53  &  6.58  &  0.011  &  ${9.5}_{-0.2}^{+0.2}$  &  126 $\pm$ 7  \\ 
ASASSN-14ae$\dagger$  &  2.61  &  5.56  &  0.013  &  ${9.5}_{-0.2}^{+0.2}$  &  41 $\pm$ 6  \\ 
PTF09axc$\dagger$  &  --  &  --  &  --  &  ${9.2}_{-0.2}^{+0.2}$  &  60 $\pm$ 4  \\ 
PTF09djl$\dagger$  &  --  &  --  &  --  &  ${9.1}_{-0.3}^{+0.3}$  &  64 $\pm$ 7  \\ 
PTF-09ge$\dagger$  &  4.03  &  6.26  &  0.019  &  ${9.2}_{-0.3}^{+0.2}$  &  59 $\pm$ 9  \\ 
PS1-10jh$\dagger$  &  --  &  --  &  --  &  ${8.7}_{-0.4}^{+0.3}$  &  65 $\pm$ 3  \\ 
iPTF15af$\dagger$  &  3.54  &  7.29  &  0.01  &  ${9.5}_{-0.2}^{+0.2}$  &  98 $\pm$ 11  \\ 
iPTF16axa  &  --  &  --  &  --  &  ${9.4}_{-0.2}^{+0.2}$  &  82 $\pm$ 3  \\ 
AT2018dyk$\dagger$  &  2.48  &  7.74  &  0.034  &  --  &  --  \\ 
ASASSN18zj$\dagger$  &  1.98  &  5.92  &  0.009  &  --  &  --  \\ 

%% file: tde_table_bpt.txt
\\[-\normalbaselineskip]
\multicolumn{7}{c}{X-ray TDEs} \\
\hline
ASASSN14li$\dagger$  &  47.7  &  22.8  &  42.5  &  17.7  &  89.3  &  AGN/Seyfert  &  0.01$\pm$0.04  &  GALAXY              BROADLINE              \\ 
Swift J1644  &  --  &  --  &  --  &  --  &  --  &  --  &  -  &                                                      \\ 
XMM J0740  &  --  &  --  &  --  &  --  &  --  &  --  &  0.01$\pm$0.03  &                                                      \\ 
ASASSN15oi$\dagger$  &  --  &  --  &  --  &  --  &  --  &  --  &  0.06$\pm$0.06  &                                                      \\ 
\hline
\multicolumn{7}{c}{Likely X-ray TDEs} \\
\hline
SDSS J1201  &  --  &  --  &  --  &  --  &  --  &  --  &  0.22$\pm$0.07  &                                                      \\ 
2MASX J0249  &  40.5  &  11.0  &  10.9  &  8.7  &  27.1  &  composite  &  0.05$\pm$0.04  &                                                      \\ 
PTF10iya  &  --  &  --  &  --  &  --  &  --  &  --  &  0.73$\pm$0.14  &                                                      \\ 
SDSS J1311  &  --  &  --  &  --  &  --  &  --  &  --  &  -  &                                                      \\ 
SDSS J1323  &  2.5  &  4.8  &  --  &  5.3  &  0.6  &  --  &  0.08$\pm$0.06  &  QSO                                        \\ 
3XMM J1521  &  --  &  --  &  --  &  --  &  --  &  --  &  0.07$\pm$0.26  &                                                      \\ 
3XMM J1500  &  --  &  --  &  --  &  --  &  --  &  --  &  0.37$\pm$0.46  &                                                      \\ 
PS18kh$\dagger$  &  --  &  --  &  --  &  --  &  --  &  --  &  -0.13$\pm$0.12  &                                                      \\ 
\hline
\multicolumn{7}{c}{Possible X-ray TDEs} \\
\hline
RX J1242-A  &  --  &  --  &  --  &  --  &  --  &  --  &  -0.03$\pm$0.04  &                                                      \\ 
RX J1242-B  &  --  &  --  &  --  &  --  &  --  &  --  &  -  &                                                      \\ 
RX J1420-A  &  5.0  &  2.5  &  1.2  &  5.1  &  0.4  &  SF  &  0.12$\pm$0.06  &  ROSAT D                                    \\ 
RX J1420-B  &  87.0  &  64.7  &  13.2  &  203.7  &  45.6  &  SF  &  -  &  ROSAT D                                    \\ 
SDSS J0159  &  30.5  &  6.5  &  19.2  &  12.4  &  13.4  &  AGN/Seyfert  &  0.53$\pm$0.07  &  QSO                 STARBURST BROADLINE    \\ 
RBS 1032  &  6.7  &  2.3  &  3.6  &  1.7  &  6.3  &  AGN/Seyfert  &  0.09$\pm$0.05  &  GALAXY                                     \\ 
RX J1624  &  --  &  --  &  --  &  --  &  --  &  --  &  0.06$\pm$0.05  &                                                      \\ 
NGC 5905  &  23.6  &  6.4  &  13.5  &  6.4  &  2.9  &  composite  &  0.15$\pm$0.03  &                                                      \\ 
\hline
\multicolumn{7}{c}{Optical/UV TDEs} \\
\hline
iPTF16fnl$\dagger$  &  --  &  --  &  --  &  --  &  --  &  --  &  0.01$\pm$0.05  &                                                      \\ 
PS16dtm$^d$  &  287.3  &  82.5  &  70.9  &  82.5  &  166.4  &  SF  &  0.44$\pm$0.07  &  GALAXY              STARFORMING            \\ 
F01004  &  --  &  --  &  --  &  --  &  --  &  --  &  \textbf{1.63$\pm$0.03}  &                                                      \\ 
GALEX D23H-1  &  21.2  &  5.9  &  15.2  &  7.6  &  2.1  &  composite  &  0.36$\pm$0.13  &                                                      \\ 
PS1-11af  &  --  &  --  &  --  &  --  &  --  &  --  &  -  &                                                      \\ 
SDSS J0952$^*$  &  620.7  &  188.4  &  580.4  &  77.9  &  163.4  &  composite  &  \textbf{1.02$\pm$0.04}  &  QSO                 BROADLINE              \\ 
SDSS J1342$^*$  &  311.0  &  70.6  &  62.3  &  55.4  &  82.8  &  SF  &  \textbf{1.08$\pm$0.03}  &  GALAXY              STARFORMING            \\ 
SDSS J1350$^*$  &  202.7  &  48.4  &  193.1  &  36.0  &  49.3  &  AGN/LINER  &  \textbf{0.87$\pm$0.03}  &  GALAXY              BROADLINE              \\ 
SDSS TDE1  &  --  &  --  &  --  &  --  &  --  &  --  &  0.2$\pm$0.24  &                                                      \\ 
SDSS TDE2$\dagger$  &  --  &  --  &  --  &  --  &  --  &  --  &  0.52$\pm$0.14  &                                                      \\ 
SDSS J0748$\dagger$$^*$  &  149.6  &  51.6  &  55.9  &  46.8  &  11.4  &  SF  &  0.77$\pm$0.03  &  GALAXY              STARFORMING            \\ 
ASASSN14ae$\dagger$  &  13.2  &  6.0  &  7.3  &  1.9  &  37.8  &  AGN/Seyfert  &  0.15$\pm$0.05  &  GALAXY                                     \\ 
PTF09axc$\dagger$  &  --  &  --  &  --  &  --  &  --  &  --  &  0.15$\pm$0.08  &                                                      \\ 
PTF09djl$\dagger$  &  --  &  --  &  --  &  --  &  --  &  --  &  0.33$\pm$0.56  &                                                      \\ 
PTF09ge$\dagger$  &  17.8  &  1.8  &  16.8  &  13.1  &  12.1  &  AGN/Seyfert  &  0.13$\pm$0.04  &  GALAXY                                     \\ 
PS1-10jh$\dagger$  &  --  &  --  &  --  &  --  &  --  &  --  &  $<$0.88   &                                                      \\ 
iPTF15af$\dagger$  &  9.0  &  --  &  --  &  --  &  2.4  &  --  &  0.18$\pm$0.08  &  GALAXY                                     \\ 
iPTF16axa  &  --  &  --  &  --  &  --  &  --  &  --  &  0.29$\pm$0.18  &                                                      \\ 
AT2018dyk$\dagger$  &  120.4  &  20.3  &  182.2  &  98.4  &  67.6  &  AGN/LINER  &  0.03$\pm$0.03  &  GALAXY                                     \\ 
AT2018bsi$\dagger$  &  --  &  --  &  --  &  --  &  --  &  --  &  0.09$\pm$0.05  &                                                      \\ 
ASASSN18zj$\dagger$  &  9.3  &  7.9  &  5.4  &  --  &  8.7  &  composite  &  0.15$\pm$0.05  &  GALAXY                                     \\ 